# Mapping generative AI use in the human brain: divergent neural, academic, and mental-health profiles of functional versus socio-emotional AI use

Junjie Wang [1,2,3], Xianyang Gan [1,2,3], Dan Liu [1,2,3], Jingxian He [1,2,3], Stefania Ferraro [1,2,3], Keith M. Kendrick [1,2,3], Weihua Zhao [1,2,3], Shuxia Yao [1,2,3], Christian Montag [4,5,6], Benjamin Becker [7,8*]

[1] The Center of Psychosomatic Medicine, Sichuan Provincial Center for Mental Health, Sichuan Provincial People's Hospital, University of Electronic Science and Technology of China, Chengdu, China

[2] The Clinical Hospital of Chengdu Brain Science Institute, MOE Key Laboratory for Neuroinformation, School of Life Science and Technology, University of Electronic Science and Technology of China, Chengdu, China

[3] Brain-Computer Interface & Brain-Inspired Intelligence Key Laboratory of Sichuan Province, University of Electronic Science and Technology of China, Chengdu, China

[4] Centre for Cognitive and Brain Sciences, Institute of Collaborative Innovation, University of Macau, Macao SAR, China

[5] Department of Psychology, Faculty of Social Sciences, University of Macau, Macao SAR, China

[6] Department of Computer and Information Science, Faculty of Science and Technology, University of Macau, Macao SAR, China

[7] MIND & AI Lab, Department of Psychology, The University of Hong Kong, Hong Kong SAR, China

[8] SRT AI, Society & Social Dynamics, Faculty of Social Sciences, The University of Hong Kong, Hong Kong SAR, China

*Corresponding author: Benjamin Becker
Department of Psychology, The University of Hong Kong, Hong Kong SAR, 999077, China.
Email address: ben_becker@gmx.de (B. Becker).

**Abstract**

The widespread adoption of generative artificial-intelligence conversational agents (AICAs) among university students constitutes a novel cognitive-social environment whose impact on the maturing brain remains elusive. Combining surveys with high-resolution structural MRI, we examined patterns of general, functional, and socio-emotional AICA use, academic performance, mental health, and brain structural signatures in a comparatively large sample of 222 young individuals. Across computational-anatomy, meta-analytic network-level, and behavioral decoding analyses, we observed use-specific associations. Higher general and functional AICA use frequencies were linked to better academic outcomes (GPA), larger dorsolateral prefrontal and calcarine gray matter volume, and enhanced hippocampal network clustering and local efficiency. In contrast, more frequent socio-emotional AICA use was associated with poorer mental health (depression, social anxiety) and lower volume of superior temporal and amygdalar regions central to social and affective processing. These findings indicate that the same class of AI tools exerts distinct effects depending on usage patterns and motivations, engaging prefrontal-hippocampal systems that support cognition versus socio-emotional systems that may track distress-linked usage. These heterogeneities are crucial for designing environments that harness the educational benefits of AI while mitigating mental-health risks.

*Keywords:* artificial intelligence; artificial intelligence-based conversational agent; human-AI interaction; chatbot; AI agent; sMRI; gray matter; Brain; depression, prefrontal cortex, amygdala, hippocampus

# Introduction

Artificial intelligence (AI) tools, in particular AI-driven conversational agents (AICAs) such as ChatGPT, Gemini, and DeepSeek, are being adopted at an unprecedented pace and scale, with over 900 million weekly users of ChatGPT alone (OpenAI, Scaling AI for everyone, 2025, https://openai.com/zh-Hans-CN/index/scaling-ai-for-everyone/). Overarching conceptual frameworks propose that these technologies may profoundly influence human cognition, emotion, social bonding, and mental well-being [1–3], yet the rapid development has outpaced empirical investigation [4,5].

Nearly half of all users (46%) are aged 18-25, and these younger users rely on AICAs more heavily for educational and personal purposes, while older users engage with them predominantly for professional tasks. University students are among the most enthusiastic early adopters, with usage rates having rocketed to 80-90% over the last year (Higher Education Policy Institute, Student Generative AI Survey 2025, 2025, https://www.hepi.ac.uk/reports/student-generative-ai-survey-2025/; Campbell University, AI in higher education: a summary of recent surveys of students and faculty, 2025,https://sites.campbell.edu/academictechnology/2025/03/06/ai-in-higher-education-a-summary-of-recent-surveys-of-students-and-faculty/), driven by AICAs' capacity to offer academic and emotional support during this uniquely formative and stressful developmental period (OpenAI, Scaling AI for everyone, 2025, https://cdn.openai.com/global-affairs/openai-edu-ai-ready-workforce.pdf).

The university years coincide with a critical developmental window in which socio-emotional capacities, executive functions, and social identity are consolidated, alongside foundational competencies that shape career trajectories and long-term personality development. The transitions during the university years unfold during the final stages of adolescent brain reorganization, extending into the mid-twenties, when prefrontal systems for executive control, emotion regulation, and social cognition are still maturing and remain highly sensitive to environmental input [6]. From a neurodevelopmental perspective, intensive AICA use in this period constitutes a novel and pervasive cognitive-social environment whose impact on neural systems for learning, executive control, and socio-affective functioning remains largely unknown, yet is potentially far-reaching.

Initial evidence and public debates on the effects of prolonged use of AICAs in young people remain deeply divided. On the one hand, studies in educational and organizational settings report that AICA engagement can enhance academic performance, higher-order cognition, and problem solving [7–9] and provide efficient socio-emotional support [10,11]. On the other hand, theoretical frameworks and emerging empirical findings highlight potential adverse consequences for executive functioning and learning [3,12] as well as decreased cognitive depths and neural engagement during learning [13]. These findings raise concerns about performance costs once AI support is removed, suggesting that AICAs may often be used as outcome-focused cognitive shortcuts that, in the long run, could compromise the development of underlying cognitive functions [14]. Frequent AICA use has also been associated with stronger instrumental and socio-emotional dependence as well as increased loneliness or distress in vulnerable users [15,16].

These concerns are partly supported by a rapidly increasing body of work over the last 15 years showing that digital technologies such as search engines, smartphones, and social media can alter executive functions and learning [e.g., 17–20] and exhibit a bidirectional relationship with mental health outcomes, in particular depression and anxiety [e.g., 21,22]. A growing literature further links prolonged engagement with these technologies to the structural and functional architecture of the brain [23,24], and initial randomized longitudinal studies indicate that sustained engagement can induce neuroplastic adaptations in systems involved in executive control, learning, and goal-directed behavior [e.g., 25–27].

Compared with earlier digital technologies, AICAs represent a qualitatively new class of tools in terms of human-technology interaction, enabling delegation of complex cognitive operations and emotionally meaningful exchanges, and they have been adopted especially rapidly and widely by young people. However, the impact of such frequent interactions on the architecture of the human brain has not yet been systematically examined.

The consequences of human-AICA interaction are also likely to vary across individuals and usage contexts. Individuals with higher levels of impulsivity, loneliness, or social anxiety, as well as related brain morphological bases, may be at greater risk of frequent or even problematic AICA use [28–32]. In addition, behavioral and neural effects may differ substantially depending on usage patterns, particularly across two broad modes of use: (i) functional, epistemic use aimed at information acquisition, problem solving, and instrumental support, and (ii) socio-emotional use centered on companionship, emotion regulation, and relational or parasocial bonding [16,28,33,34]. Initial work suggests that these patterns dissociate behaviorally—for example, functional use has been linked to job insecurity, whereas socio-emotional use relates more strongly to problematic use [16,29,33]—raising the possibility that they also map onto distinct behavioral, mental health and brain structural profiles.

Here we present the first systematic characterization of the neuroanatomical correlates of general as well as specific AICA use in relation to educational outcome and mental health indices. We combined high-resolution T1-weighted structural magnetic resonance imaging (sMRI) in 222 students with comprehensive survey assessments and a series of computational neuroanatomical analyses to examine variations in regional morphometry and morphological networks, complemented by meta-analytic coactivation and behavioral decoding approaches to guide interpretation. This design allowed us to determine (1) associations between specific AICA use patterns (general, functional, socio-emotional) and academic and mental health outcomes; (2) associations between AICA use patterns and regional gray matter variation using computational morphometry; (3) the network-level and behavioral characteristics of the regions associated with AICA use patterns using meta-analytic connectivity modelling (MACM) and behavioral decoding analyses; and (4) topological properties of brain structural networks associated with individual usage characteristics based on graph theoretical analysis.

## Results

### Participant characteristics and AICA usage

The final sample encompassed 222 healthy adult participants (53.6% female; M ± SD = 21.32 ± 2.44; for participant dropout, see the Methods section). Most participants (58.5%) started to use AICA for the first time within about one year after the release of ChatGPT, and almost no one started using AICA during the study (0.5%; Fig. S1). All participants had used AICA recently (within one month), with 48.6% reporting AICA use within one day and 38.3% reporting AICA use within one week (Fig. S2). With respect to general AICA use frequency, the majority of participants (61.3%) used AICA frequently throughout the week (at least 4 days per week; Fig. S3). For specific AICA use frequency, the majority of participants (82.5%) reported that they frequently used AICA to meet their functional needs (Fig. S4); while a smaller number of participants (6.8%) reported that they frequently used AICA to meet their socio-affective needs (Fig. S5). And 37.5% of the participants reported that they used AICA about one hour or more per day on average (Fig. S6; further details on AICA usage see Results section and Fig. S1-15 in the supplements). General AICA use frequency was significantly correlated with functional AICA use frequency (r = 0.459, p < 0.001) but not socio-emotional AICA use frequency (r = 0.134, p = 0.050) after controlling for sex and age. The partial correlation between functional AICA use frequency and socio-emotional AICA use frequency remained insignificant (r = 0.050, p = 0.466).

**Associations with academic performance and mental health**

Examining associations between the three types of AICA use with academic performance revealed significant positive associations between grade-point average (GPA) with general AICA use frequency (r = 0.169, p = 0.013; Fig. 1a & b) and functional AICA use frequency (r = 0.182, p = 0.008, both Bonferroni corrected; Fig. 1a & c), but not socio-emotional use frequency (r = -0.065, p = 0.342; Fig. 1a). In contrast socio-emotional AICA use frequency was positively correlated with depression (r = 0.186; p = 0.006, surviving Bonferroni correction; Fig. 1a) and social anxiety (r = 0.148; p = 0.031; Fig. 1a), and was negatively correlated with general mental health (r = -0.154; p = 0.025; Fig. 1a). No significant correlation was observed between general AICA use frequency and mental health, whereas functional AICA use frequency showed a positive correlation with social anxiety (r = 0.142, p = 0.039; Fig. 1a). Details on age, total intracranial volume (TIV), general AICA use frequency, functional AICA use frequency, and socio-emotional AICA use frequency in the total sample, female sample, and male sample were shown in the Table 1, and GPA and mental health variables were shown in the Table S1.

Table 1. Participant information

| Variable | Overall (n = 222) | | Male (n = 103) | | Female (n = 119) | |
|---|---|---|---|---|---|---|
| | M (SD) | Range | M (SD) | Range | M (SD) | Range |
| Age | 21.32 (2.44) | 18-32 | 21.18 (2.33) | 18-32 | 21.45 (2.52) | 18-30 |
| TIV | 1538.44 (143.52) | 1184.42-2006.16 | 1637.28 (105.48) | 1404.96-2006.16 | 1452.90 (114.24) | 1184.42-1897.60 |
| General AICA use | 6.30 (1.82) | 2-9 | 6.68 (1.72) | 3-9 | 5.97 (1.84) | 2-9 |
| Functional AICA use | 5.12 (0.86) | 2-7 | 5.14 (0.82) | 2-7 | 5.10 (0.90) | 2-7 |
| Socio-emotional AICA use | 2.22 (1.32) | 1-7 | 2.16 (1.27) | 1-7 | 2.28 (1.36) | 1-7 |

Note: AICA = Artificial Intelligence-driven Conversational Agent, M (SD) = Mean (Standard Deviation), TIV = Total Intracranial Volume.

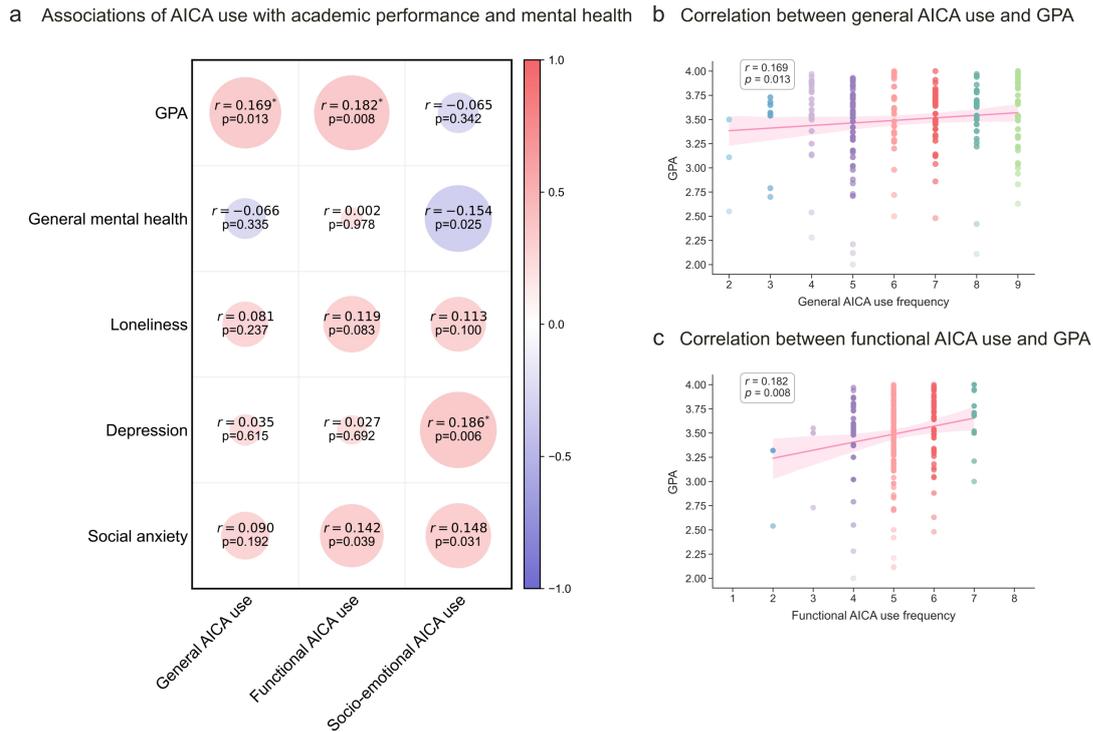

**Fig. 1. Partial correlations between AICA use frequency and academic performance and mental health, controlling for sex and age.** (a) Associations of different AICA use patterns with academic performance and mental health. Circles in red represent positive correlations, whereas circles in purple represent negative correlations. Larger and more saturated circles represent stronger correlation coefficients. The "*" sign indicates significance after Bonferroni correction at the 0.05 level. (b) Visualization of the positive association between general AICA use frequency and GPA. (c) Visualization of the positive association between functional AICA use frequency and GPA. Abbreviations: AICA = Artificial Intelligence-driven Conversational Agent, GPA = grade-point average.

**Associations between AICA use characteristics and regional brain volume**

The voxel-based morphometry (VBM) analysis showed that general AICA use frequency was positively correlated with gray matter volume (GMV) in the left dorsolateral prefrontal cortex (dlPFC), i.e., dorsolateral superior frontal gyrus (SFG; k = 1339, T = 5.27, $p_{FWE}$ = 0.009) and medial occipital lobe (calcarine fissure and cuneus, CAL; k = 1676, T = 4.28, $p_{FWE}$ = 0.003; Fig. 2a; Table S2). In the VBM analyses for specific AICA use, more frequent socio-emotional AICA use was associated with decreased GMV of the left superior temporal gyrus (STG) and amygdala (k = 1477, T = 4.35, $p_{FWE}$ = 0.006; Fig. 2b; Table S2), whereas functional AICA use frequency did not show any significant associations with regional brain volume. We additionally explored associations with the dosage of daily exposure and found that daily usage time was positively correlated with GMV in the identified dlPFC region (r = 0.161, p = 0.017, partial correlation controlling for sex, age, and TIV), underscoring exposure-related effects.

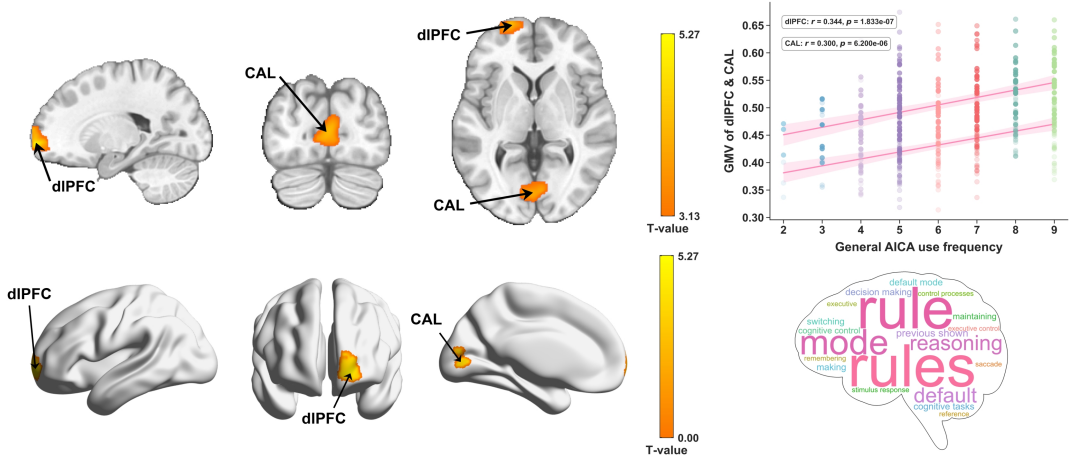

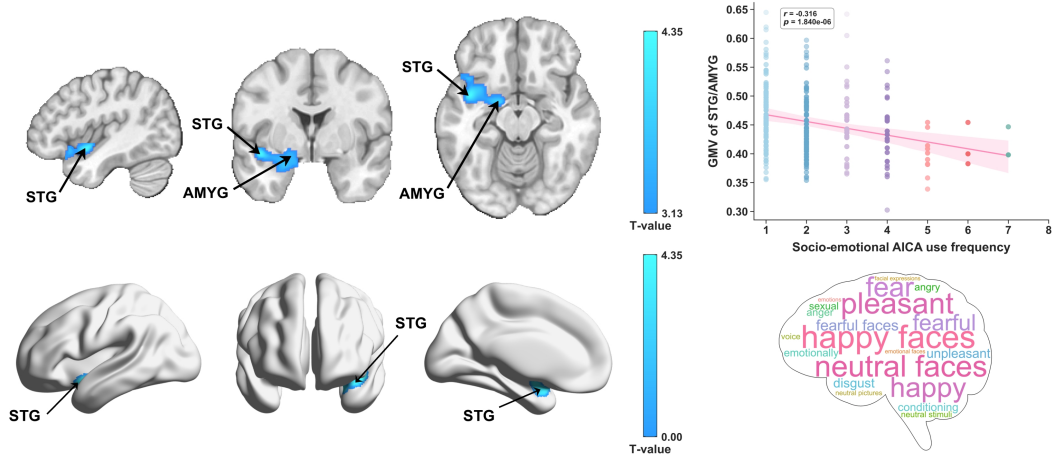

**Fig. 2. Neuroanatomical correlates of AICA use and corresponding functional annotations.** (a) Neuroanatomical correlates of general AICA use frequency and corresponding behavioral decoding. Brain maps show the regions whose GMV is correlated with general AICA use frequency in the VBM analysis; the scatter plot shows the linear relationship between regional GMV and general AICA use frequency; and the word cloud plot shows the top 20 behavioral terms related to the neural correlates of general AICA use frequency. (b) Neuroanatomical correlates of socio-emotional AICA use frequency and corresponding behavioral decoding. Brain maps show the regions whose GMV is correlated with socio-emotional AICA use frequency in the VBM analysis; the scatter plot shows the linear relationship between regional GMV and socio-emotional AICA use frequency; and the word cloud plot shows the top 20 behavioral terms related to neural correlates of socio-emotional AICA use frequency. The coordinates for the slice brain maps displayed in 1a are: x = -20, y = -81, z = 2; whereas those for 1b are: x = -44, y = 0, z = -14. Abbreviations: AICA = Artificial Intelligence-driven Conversational Agent, AMYG = Amygdala, CAL = Calcarine Fissure, dlPFC = dorsolateral Prefrontal Cortex, STG = Superior Temporal Gyrus.

**Network-level and behavioral characterization of the identified brain regions**

Subsequent analyses used meta-analytic coactivation and behavioral decoding to characterize the network and functional properties of the identified regions. The left dlPFC/SFG and left CAL, which showed larger volumes with more frequent general AICA use, coactivated with frontal and subcortical circuits implicated in cognitive and inhibitory control (left middle frontal gyrus, left inferior frontal gyrus, left thalamus, and left supplementary motor area), as well as with visual regions (right inferior occipital gyrus; Fig. 3a & b; Table S4). Behavioral decoding of the regions (Fig. 2a) and networks (Fig. 3a & b) indicated links with higher-order cognitive functions (e.g., rules, reasoning, executive functions, and decision-making), self-referential processes (e.g., default mode and reference), and visual attention.

In contrast, the left STG-amygdala cluster, which showed reduced volumes with more frequent socio-emotional AICA use, coactivated with a network of regions involved in socio-affective processing, including autonomic arousal (bilateral insula) and social processing (bilateral fusiform gyrus and left pallidum), as well as visual and somatosensory regions (for example, right inferior occipital gyrus, left precentral gyrus, and right supplementary motor area; Fig. 3c & d; Table S4). Behavioral decoding linked the regions (Fig. 2b) and this network (Fig. 3c & d) to social and affective processing (for example, happy and neutral faces, pleasantness, happiness, fear). A detailed description of the coactivation patterns is available in the Table S4 and Results section of the Supplementary Material.

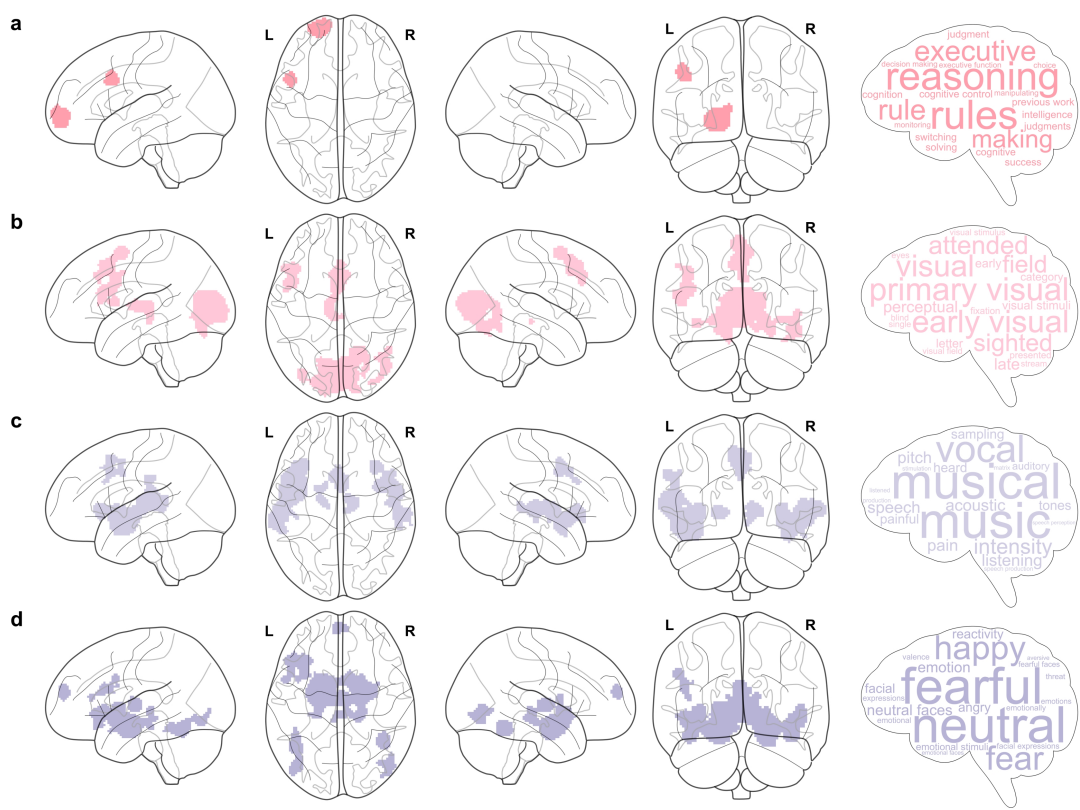

**Fig. 3. Functional coactivation networks of brain regions associated with AICA use and corresponding behavioral characterization.** The brain maps in Fig. 3a-d separately show the coactivation patterns of the left dlPFC, CAL, STG, and amygdala revealed in the VBM analysis. The word cloud maps in Fig. 3a-d separately display the top 20 behavioral terms that are most strongly correlated with corresponding coactivation patterns, with a larger and more opaque font size reflecting a stronger correlation. Abbreviations: L = Left Hemisphere, R = Right Hemisphere.

## Associations between AICA use and brain network properties

The graph theoretical brain structural network analyses revealed a positive correlation between general AICA use frequency and clustering coefficient of the right hippocampus (r = 0.277, p$_{FDR}$ = 0.003; Fig. 4a), a negative correlation between socio-emotional AICA use frequency and clustering coefficient of the left cuneus (r = -0.232, p$_{FDR}$ = 0.048; Fig. 4b), and a positive correlation between general AICA use frequency and the local efficiency of the right hippocampus (r = 0.250, p$_{FDR}$ = 0.016; Fig. 4c). No other significant associations between AICA use frequency and nodal or global graph metrics were detected (p$_{FDR}$ > 0.05).

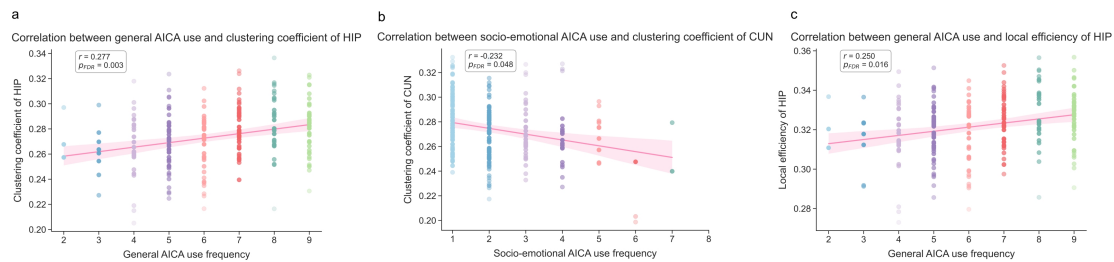

**Fig. 4. Associations between AICA use and brain morphological network properties.** (a) Correlation between general AICA use frequency and clustering coefficient of HIP. (b) Correlation between socio-emotional AICA use frequency and clustering coefficient of CUN. (c) Correlation between general AICA use frequency and local efficiency of HIP. Abbreviations: AICA = Artificial Intelligence-driven Conversational Agent, FDR = False Discovery Rate, CUN = Cuneus, HIP = Hippocampus.

## Discussion

Our study provides the first systematic, population-based evidence that frequent use of AICAs in young adults is associated with distinct, use-specific variations in brain structure at both regional and network levels. In addition, the findings extend recent work by delineating how different AICA use patterns relate to academic performance and mental health outcomes. In our university sample, functional AICA use was widespread, with more than 80% of students reporting frequent use, whereas socio-emotional use was less common but still substantial, with around 7% using AICAs frequently for socio-emotional support. General and functional AICA use frequencies were strongly positively correlated and were associated with better academic outcomes and higher GMV in prefrontal (dlPFC) and occipital (CAL and cuneus) regions related to executive function networks and behavioral domains. By contrast, more frequent socio-emotional AICA use was associated with lower volume in the left STG and amygdala, regions embedded in affective networks and characterized by socio-emotional processing. Complex network analyses underscored the distinct brain structural associations on the network-level such that more frequent general AICA use was linked to greater clustering coefficient and local efficiency of the right hippocampus, whereas more frequent socio-emotional AICA use was linked to reduced clustering coefficient in the medial occipital lobe (i.e., cuneus).

In line with reports documenting AICA usage rates of 80–90% among Western university students (Higher Education Policy Institute, Student Generative AI Survey 2025, 2025, https://www.hepi.ac.uk/reports/student-generative-ai-survey-2025/; Campbell University, AI in higher education: a summary of recent surveys of students and faculty, 2025, https://sites.campbell.edu/academictechnology/2025/03/06/ai-in-higher-education-a-summary-of-recent-surveys-of-students-and-faculty/), our data indicate that more than 80% of students in our Chinese sample are current, frequent AICA users. The general and functional uses of AICA—but not socio-emotional use—were moderately to strongly correlated (r = 0.459, p < 0.001), suggesting that most students engage with AICAs primarily for instrumental, task-oriented support in their studies and academic work. Within our sample, more frequent AICA use was associated with higher GPA, aligning with recent meta-analyses and empirical studies reporting beneficial effects of AICA use on academic performance [e.g., 7–9].

We further demonstrate for the first time that more frequent general AICA use is associated with region-specific volumetric increases in the left dlPFC and medial occipital lobe, alongside greater clustering coefficient and local efficiency of the right hippocampus. Functional network-level and behavioral decoding analyses indicate that these regions form part of the frontal-thalamic circuits implicated in cognitive and inhibitory control [e.g., 35] and map onto behavioral domains related to higher-order cognitive and self-referential functions.

Previous work has linked prolonged engagement with new technologies, such as video games and social media, to alterations in brain structure, predominantly on the basis of cross-sectional studies with limited causal conclusions [e.g., 36,37]. However, initial randomized controlled trials provide converging evidence that cognitively demanding digital engagement can induce structural plasticity: two months of video game training increased GMV in the dlPFC and hippocampus and improved cognitive performance, whereas six weeks of intensive online gaming reduced parts of the orbitofrontal GMV in the context of increased dependency [25,27]. Together, these findings suggest that sustained interaction with novel technologies can give rise to regional neuroplastic adaptations that depend on the specific cognitive and affective demands involved. The supply-demand model of neurocognitive plasticity proposes that cognitive and brain structural change are driven by long-term mismatches between capacity and environmental demands [38], analogous to neuroplasticity induced by learning [6,39,40]. During frequent functional interactions with AICAs, the human brain must continually adapt to the cognitive demands of the interactive environment, including precise linguistic formulation and logical rigor in prompt engineering [41,42], rapid assimilation of high-quality information [43], and ongoing critical evaluation and value assessment of AI-generated content [44,45]. In the context of AICAs, transformed modes of cognitive engagement and para-social interaction may similarly create such imbalances, leading to use-pattern-specific neurostructural adaptations [1,3].

In the present study, more frequent AICA use was associated with larger dlPFC volume, a region identified as a central hub for core executive functions [46–48] and metacognitive capacity [49,50], with larger dlPFC/SFG gray matter typically linked to superior executive functioning [51,52] and metacognitive ability [53,54]. In addition, functional engagement with AICAs relies on reading, visual comprehension, and learning, processes that have been associated with demand-specific volumetric increases in the medial occipital lobe, including the CAL and cuneus [55,56]. The hippocampus is critical for learning and cognition [57–59], with larger hippocampal volumes linked to better performance on executive function tasks [60] and longitudinal studies reporting hippocampal gray matter increases following intensive periods of learning [61]. It is therefore plausible that these regions are heavily recruited to meet the high cognitive resource demands of human-AI interactions, undergoing adaptive changes during prolonged, high-frequency cognitive engagement. Converging with this mechanistic interpretation, a growing body of work suggests that AICA use can bolster higher-order cognitive functions [8,44,62–65], with the present data pointing to corresponding neuroplastic adaptations.

Beyond plasticity, these neural correlates may also reflect pre-existing neurostructural predispositions that make some individuals more likely to engage frequently with AICAs for functional and academic purposes. Prior work links the dlPFC circuitry to motivated, goal-directed behavior [66] and curiosity-driven learning [67] and shows that prefrontal systems in concert shape persistence toward long-term goals [68]. In the context of goal-oriented AICA use and the academic context, such traits could predispose individuals to adopt more intensive functional usage patterns.

A substantial subgroup in our sample reported frequent socio-emotional use of AICAs, and higher engagement in this mode was associated with elevated depression and social anxiety, as well as poorer self-reported general mental health. While initial randomized trials suggest that purpose-built therapeutic AICAs can alleviate depressive and anxiety symptoms [e.g., 69,70], emerging observational evidence indicates that frequent socio-emotional use of general AICAs in everyday contexts may be linked to higher levels of depression and anxiety [71,72]. Additional work suggests that lower social competence can predispose individuals to more sustained AI engagement [16], and that greater psychological distress may motivate the use of AICAs for socio-emotional support [73], consistent with broader models proposing that internalizing psychopathology can drive problematic technology use [74]. Our behavioral findings align with this broader literature, showing that more frequent socio-emotional AICA use is significantly associated with poorer mental health outcomes.

At the neural level, higher socio-emotional AICA engagement was linked to reduced GMV in the STG and amygdala, regions that play central roles in social information processing [75–78], with the amygdala serving as a core hub for affective arousal and fear [79,80]. Gray matter reductions in these regions have been extensively documented in internalizing disorders, including anxiety [18,81–86] and depression [87–91]. Longitudinal studies further indicate that smaller amygdalar volumes can precede and predict later psychopathology and dysfunctional behavior, including aggression [92] and impaired emotion regulation [93–95] and cross-sectional work consistently links lower amygdala volume to higher anxiety, depression, neuroticism, and alexithymia [96–101]. In the context of these behavioral and neuroimaging lines of evidence, our findings suggest that the neuroanatomical correlates of socio-emotional AICA use may, at least in part, reflect higher levels of internalizing psychopathology and reduced temporal and amygdalar volumes that predispose some individuals to seek socio-emotional support from AICAs. This interpretation is further supported by the meta-analytic functional network and behavioral characterizations of the regions and networks indicating an involvement in socio-emotional processes.

Although frequent socio-emotional use is confined to a small but meaningful subset of users—consistent with recent large-scale data [16]—and most users primarily seek functional support, a sizeable proportion of the overall user base may nevertheless be at risk of engaging in intensive socio-emotional interactions with AICAs. This underscores the urgent need to systematically monitor the quality, safety, and reliability of socio-emotional guidance provided by these systems [102].

While the present study offers the first characterization of use-mode-specific associations between AICA engagement and brain structure, these findings should be interpreted in light of several limitations. Future work will need (1) replication in prospective longitudinal designs to establish the directionality of effects, (2) detailed examination of individual differences beyond psychopathological vulnerability, including age, gender, and academic major, and (3) systematic investigation of the acute and long-term cognitive and brain functional consequences of AICA use. Further, we speak in this work in general about AICA usage, but it is well-known that certain products such as Replika are more specifically used for socio-emotional interaction. Hence, taken a deeper look at platform-dependent usage presents an important future research endeavor (which is also a lesson learned from social media research with different platforms exerting different addictive behaviors) [103].

Our study provides the first systematic evidence that everyday AICA use in young adults relates to distinct neurostructural profiles and behavioral patterns, differentiating a predominantly functional, academically oriented usage mode from a socio-emotional, distress-linked mode. Frequent general AICA use was associated with greater integrity of prefrontal-occipital-hippocampal systems and better academic outcomes, whereas frequent socio-emotional use was linked to reduced temporal-amygdalar volume and poorer mental health. Together, these findings outline a neurobiological framework for how generative AI tools intersect with human brain, suggesting that they may both scaffold cognitive development in some users and interact with pre-existing vulnerabilities that predispose others to intensive, distress-driven socio-emotional engagement.

## Methods

### Participants

An initial sample containing 248 young and healthy adults was collected for exploring the neuroanatomical correlates of AICA use. However, several participants were excluded from the analysis due to poor quality of sMRI data after visual inspection (n = 2) and poor quality of self-reported behavioral data (n = 8). In addition, the participants who reported no AICA use within the last month (n = 16) were treated as non-AICA user and then excluded, resulting that a final clean sample of AICA users (n = 222; 119 females) was included in the analysis. No participant reported a history of, or current physical disease and mental disorder. All participants signed the informed consent form and completed the survey after sMRI scanning. The protocol was approved by the ethics committee of University of Electronic Science and Technology of China and followed the latest version of Declaration of Helsinki.

### Measures

**AICA use frequency.** Based on prior studies [33,72,104–106], AICA usage was assessed in three domains: 1) general AICA use frequency ("How often do you use AICA on average per week?", 9-point Likert scale ranging from 1 [never used] to 9 [7 days a week]); 2) functional AICA use frequency ("To what extent do you use AICA to obtain functional support, such as advice, opinion, and assistance for work, study, and academic?", 7-point Likert scale ranging from 1 [never used] to 7 [extremely frequently use]); 3) socio-emotional AICA use ("To what extent do you use AICA to obtain socio-emotional support, such as confiding in AI, receiving AI companionship, and developing friendship or romantic relationship with AI?", 7-point Likert scale ranging from 1 [never used] to 7 [extremely frequently use]; see details in the Supplementary Material). Items were adapted from previous studies, with higher scores reflecting more frequent use [72,104–106]. In addition to use frequency, we designed a series of items—including such as the "last time of using AICA" (to exclude non-AICA users) and "AICA daily usage time" (for validation of brain-AICA use associations)—to determine the AICA usage profile of the student population (item descriptions are detailed in the appendix section of the Supplementary Material).

**Academic performance, mental health, and demographics.** To explore relationships between AICA use and academic performance, the GPA of each participant was collected as a measure of academic outcome. Each participant was required to look up their GPA in the university's academic system and fill in it (rounded to two decimal places). For mental health, scales and items were included to assess general mental health status, loneliness, depression, social anxiety over the past month. Depression was measured by one item from the Screening Tool for Psychological Distress, which has been proved to be efficient and highly consistent with the Beck Depression and Anxiety Inventory [107,108]. Participants rated the item on a 9-point Likert scale ranging from 1 (not at all) to 9 (severely), with a higher score signifying a severer depression. Social anxiety was measured by the 10-item version Social Interaction Anxiety Scale (SIAS-10), which was widely used in numerous studies and shown to possess considerable reliability [109,110]. Participants were asked to select the option that best described themselves on a 5-point Likert scale ranging from 1 (not at all characteristic or true of me) to 5 (extremely characteristic or true of me). The total scores of 10 items were calculated, with a range from 10 to 50, and a higher score denoted a severer social anxiety. In our sample, the Cronbach's alpha was 0.926 for the SIAS-10, existing a high internal consistency. General mental health status and loneliness over the last month were measured by single self-designed items (please see full descriptions in the Supplementary Material). For demographical data, sex and age were collected and served as control variables in the subsequent analyses.

**MRI data acquisition and preprocessing**

All participants underwent high-resolution brain structural scanning on a 3.0 T GE MR750 system. Whole-brain T1-weighted images were acquired using a 3D spoiled gradient echo pulse sequence. After visual inspection, the sMRI data were processed using Computational Anatomy Toolbox (CAT12, r2577; https://neuro-jena.github.io/cat) with a standard pipeline [111–113]. Then quality checks were performed to exclude any participant with low image quality and TIV was estimated to show and control individual differences on brain size [114]. Detailed information about acquisition parameter, preprocessing pipeline, and image quality is available in the Supplementary Material.

**Regional morphology analysis**

The modulated and smoothed gray matter maps were subjected to a multiple linear regression model in the Statistical Parametric Mapping software package (SPM12; https://www.fil.ion.ucl.ac.uk/spm/software/spm12/) to determine associations between GMV variations and general AICA use, functional AICA use, and socio-emotional AICA use, respectively [115], with sex, age, and TIV as covariates. We applied a mask with an absolute threshold of 0.1 to exclude voxels with low gray matter probability and control noises as well as adopted an implicit brain mask implemented in CAT12 to ensure valid data across all subjects (recommended parameters in the CAT12 manual; https://neuro-jena.github.io/cat12-help/#intro). An uncorrected voxel-level threshold of $p < 0.001$ and a family-wise error (FWE)-corrected cluster-level threshold of $p < 0.05$ were applied on the statistical image to examine brain clusters significantly correlated with AICA use [116,117]. Finally, anatomical labeling was performed by mapping the peak coordinate within each cluster to the Automated Anatomical Labeling (AAL) atlas [118]. In cases where a coordinate fell outside the boundaries of all defined AAL regions, the label of the nearest AAL region was assigned. We extracted the mean GMV of the significant clusters and then examined Pearson partial correlation coefficient between mean GMV and corresponding AICA use frequency, with sex, age, and TIV as covariates. Additionally, we also conducted similar VBM analyses on the sample including participants with infrequent AICA use (n = 238) and the results largely replicated the findings above (Fig. S16; Table S3).

**Functional network-level and behavior-level characterization**

**Functional network coactivation analysis.** The MACM was employed to determine which brain systems were coactivated with the regions showing GMV changes related with AICA use frequency. Following a standard MACM procedure [119], four regions of interest (ROIs) identified in the

VBM analyses were defined as spheres centered on the peak coordinates. We then retrieved dataset of each ROI used for MACM analyses from the BrainMap database using Sleuth. Finally, the coactivation pattern of each ROI was determined by an independent meta-analysis implemented by GingerALE, using a voxel-level threshold of p < 0.001 and a cluster-level FWE-corrected threshold of p < 0.05 with 5,000 permutations [120,121]. Additional methodological details and dataset characteristics are provided in the Supplementary Material.

**Behavioral decoding analysis.** To determine the cognitive functions related to the brain regions showing covariance with AICA use and their coactivation patterns, we conducted the functional annotation with the BrainStat toolbox (https://github.com/MICA-MNI/BrainStat) [128]. Consistent with prior studies [123,124], the unthresholded statistical maps exploring the correlation between GMV and general AICA use and socio-emotional AICA use (Fig. 2), and coactivation statistical maps of four ROIs showing significant covariance with AICA use (Fig. 3), were submitted to the toolbox and it automatically computed correlations between each brain map and over 3000 terms from the Neurosynth database (https://www.neurosynth.org). Then, the top 20 terms with the highest positive correlation coefficients were extracted to represent the functional characterization of the corresponding brain morphology and coactivation patterns. Of note, we excluded those non-functional terms such as visual cortex, insular cortex, mid, and matrix in the extraction.

### Morphological network-level analysis

**Morphological similarity network (MSN) construction.** The smoothed gray matter map of each subject was utilized for the individualized MSN construction with the AAL atlas [125] as the segmentation reference and Jensen-Shannon divergence as the similarity estimate method (see details in the Supplementary Material) [126]. In this way, a 90 × 90 symmetric morphological connectivity matrix with 3,960 unique morphological brain connectomes was generated for each participant, with values ranging from 0 to 1, where 0 denotes complete dissimilarity and 1 represents complete identity.

**Graph theory analysis.** For determining univariate associations between AICA use frequency and topological properties of the MSN, we calculated both global graph theory metrics (like clustering coefficient $C_p$) and nodal graph theory metrics (like nodal local efficiency $NLE_i$) using the Graph Theoretical Network Analysis Toolbox (GRETNA; available at: https://www.nitrc.org/projects/gretna) [127]. With a sparsity-based thresholding strategy, each metric was computed on each weighted network generated under different network sparsity thresholds ranging from 0.05 to 0.50 with a step size of 0.01. Then we calculated the area under the curve (AUC) for each graph metric across multiple sparsity thresholds as a comprehensive scalar [127], and correlated the AUC of each metric with AICA use frequency after controlling sex, age, and TIV. The False Discovery Rate (FDR) method was used for multiple comparison correction of several univariate correlation analyses with a significance level of p < 0.05 [128]. More methodological descriptions about the graph metrics could be found in the Supplementary Material.

## Funding sources

The present study was supported by the Ministry of Science and Technology of China (STI 2030-Major Projects 2022ZD0208500), Hong Kong University Grants Council (GRF 17615525), National Natural Science Foundation of China (NSFC 82271583), and The University of Hong Kong seed funding and start-up schemes (2407102536).

## Acknowledgments

We are grateful to Liqin Zhang, Xiangfeng Yan, Xiaodong Zhang, Luxuan Yang, Ziheng Wang, Yue Teng, Yu Wu, and Mengfan Han for their tremendous support in data collection.

## CRediT authorship contribution statement

Junjie Wang: Conceptualization, Data Curation, Formal Analysis, Investigation, Methodology, Resources, Validation, Visualization, Writing – Original Draft, Writing – Review & Editing; Xianyang Gan: Formal Analysis, Methodology, Writing – Review & Editing; Dan Liu: Data Curation, Investigation; Jingxian He: Data Curation, Investigation; Stefania Ferraro: Resources; Supervision; Writing – Review & Editing; Keith M. Kendrick: Resources; Supervision; Writing – Review & Editing; Weihua Zhao: Project Administration; Resources; Supervision; Writing – Review & Editing; Shuxia Yao: Project Administration; Resources; Supervision; Writing – Review & Editing; Christian Montag: Supervision; Writing – Review & Editing; Benjamin Becker: Conceptualization; Funding Acquisition; Methodology; Project Administration; Resources; Supervision; Writing – Review & Editing.

## Declaration of competing interest

The authors declare no conflict of interest in this study.

## Data availability

Data are available from the corresponding author upon reasonable request. The original codes of MSN construction used in this study can be found at: https://github.com/Niu619/Structure-function-coupling.

## Supplementary data

The supplementary material for this article is available online.

# Supplementary Material

## Method

**MRI data acquisition and preprocessing**

All participants underwent high-resolution brain structural scanning on a 3.0 T GE MR750 system (General Electric Medical Systems, Milwaukee, WI, USA). Whole brain T1-weighted images were collected using 3D spoiled gradient echo pulse sequence with the following acquisition parameters: voxel size = 1 mm isotropic, repetition time = 6 ms, echo time = 2 ms, flip angle = 12°, field of view = 256 × 256 mm, acquisition matrix = 256 × 256, slice thickness = 1 mm, number of slices = 156.

Visual inspection was performed on the neuroanatomical data of each participant to identify and exclude images with poor quality (e.g., severe artifacts or abnormal structure). The remaining images were processed by the Computational Anatomy Toolbox (CAT12, r2577; https://neuro-jena.github.io/cat) implemented in the Statistical Parametric Mapping software package (SPM12; https://www.fil.ion.ucl.ac.uk/spm/software/spm12/) with a recommended default pipeline [1,2]. Specifically, the preprocessing steps included bias-field correction, linear spatial normalization based on 12-parameter affine registration [3], segmentation of gray matter, white matter, and cerebrospinal fluid [4], non-linear spatial normalization to Montreal Neurological Institute (MNI) space based on the shooting algorithm [5], resampling to voxel size of 1.5 mm, and modulation of images for minimizing the distortion. A Gaussian kernel with a full-width at half-maximum of 8 mm was adopted to smooth the modulated gray matter images used for VBM (voxel-based morphology) analysis.

Finally, quality checks were performed on the automatically generated quality reports in the preprocessing. No subjects were excluded at this step due to the high image quality of all participants. Specifically, the weighted average image quality rating (IQR) of each participant was ≥ B-, representing an overall good quality judged by CAT12 and a higher quality than most open-access databases (C to C+ range). IQR was a comprehensive indicator automatically generated for the assessment of noise, inhomogeneity, and resolution in the CAT12 [2]. Additionally, the total intracranial volume was estimated as well to show and control individual differences on brain size [6].

**Functional network coactivation analysis**

The meta-analytic connectivity modelling (MACM) was employed to determine which brain systems were coactivated with the regions showing gray matter volume (GMV) changes related with artificial intelligence-based conversational agent (AICA) use frequency. MACM was developed as a robust method that provide a task-free network-level functional characterization by performing meta-analyses on studies reporting activations in region of interest (ROI) retrieved from large neuroimaging databases [7–9], and had been further demonstrated as a valid alternative to resting-state functional connectivity and white matter connectivity [10,11]. In the current study, the MACM was implemented based on a standard procedure [8]. Firstly, we defined four ROIs for the left dorsolateral prefrontal cortex (dlPFC): -20, 66, -4; left calcarine fissure and surrounding cortex (CAL): -2, -81, 9; left superior temporal gyrus (STG): -44, 2, -12; left amygdala: -18, -3, -14, which were revealed in the VBM analyses (Fig. 2 and Table S2). Concretely speaking, we drew a sphere centered on the peak coordinate of left amygdala with a radius of 6 mm, and drew spheres separately centered on the peak coordinate of the other 3 brain cortical regions with a radius of 8 mm. Secondly, we set a series of commonly used filters ("Activations: Activations only", "Diagnosis: Normals", "Imaging Modality:

fMRI", and "Context: Normal Mapping") in Sleuth 3.0.4 (https://www.brainmap.org/sleuth/) to obtain study datasets from the Brainmap database [12]. After meticulous screening, datasets included in the MACM analyses of 5 ROIs were in order as follows: 1) 27 experiments, 474 foci, and 378 subjects; 2) 137 experiments, 2170 foci, and 1894 subjects; 3) 83 experiments, 1157 foci, and 1325 subjects; 4) 145 experiments, 1880 foci, and 2352 subjects. Thirdly, all coordinates in four datasets were exported separately to a txt file with experiment as the organization criteria and Montreal Neurological Institute as standard space. Fourthly, we performed the meta-analysis for each ROI in the GingerALE 3.0.2 (https://www.brainmap.org/ale) and identified the co-activation pattern of each ROI by applying an uncorrected voxel-level threshold of p < 0.001 and an family-wise error (FWE)-corrected cluster-level threshold of p < 0.05 with 5,000 permutations [13–15]. The non-ROI regions activated in the meta-analysis were recognized as co-activated regions with the ROI.

**Morphological similarity network construction**

The smoothed gray matter (GM) map of each subject was utilized for the individualized morphological similarity network (MSN) construction with a Jensen-Shannon Divergence (JSD) method [16]. As a symmetric variant of the Kullback-Leibler Divergence (KLD) with finite values, it addressed the mathematical limitations of KLD such as asymmetry and unboundedness [17], had been proven superior to KLD [16], and had been successfully applied to portray morphological network alterations regarding aging and brain disorders [18–20]. To be specific, GM maps were firstly segmented as 90 brain regions predefined by AAL atlas, a commonly used parcellation in prior MSN studies [21,22]. Then, the probability density estimate of each brain region was calculated by a kernel density estimation (a function implemented in Matlab, available at: https://www.mathworks.com/matlabcentral/fileexchange/14034-kernel-density-estimator) with a sampling resolution of 27 (i.e., 128) points [23,24], and was converted into the probability density functions (PDFs). Subsequently, the morphological brain connectome between each pair of brain regions was determined by computing statistical similarity of PDFs between two regions with JSD method [16]. Finally, a 90 × 90 symmetric morphological connectivity matrix with 3,960 unique morphological brain connectomes was generated for each participant, with values ranging from 0 to 1, where 0 denotes complete dissimilarity and 1 represents complete identical.

**Graph theory analysis**

For fully determining univariate associations between AICA use frequency and topological properties of MSN, we calculated both global graph theory metrics (hierarchy $\beta$, synchronization $S$, assortativity $r$, global efficiency $E_{glob}$, local efficiency $E_{loc}$, small-worldness $\sigma$, clustering coefficient $C_p$, standardized clustering coefficient $\gamma$, shortest path length $L_p$, and standardized shortest path length $\lambda$) and nodal graph theory metrics (nodal degree $K_i$, nodal betweenness $B_i$, nodal efficiency $E_i$, nodal local efficiency $NLE_i$, and nodal shortest path $NLP_i$) using the Graph Theoretical Network Analysis Toolbox (GRETNA; available at: https://www.nitrc.org/projects/gretna) [25]. The definition, explanation, and computational formula of these metrics can be found in the prior studies [26,27]. To exclude the influence of overall connectivity strength on topological metrics, we employed a sparsity-based thresholding strategy. The sparsity range was determined based on the following criteria: 1) the averaged degree over all nodes of each network was larger than 2 × log(N), where N is the number of nodes; 2) the small-worldness of the networks was larger than 1.1 for all subjects, ensuring the small-worldness and sparsity of networks [28]. Hence, each metric was computed on each weighted network generated under different network sparsity thresholds ranging from 0.05 to 0.50 with a step size of 0.01, and then each metric represented a function of sparsity. With this in mind, we calculated the area under the curve (AUC) for each graph metric across multiple sparsity thresholds as a comprehensive scalar [25], and correlated the AUC of each metric with AICA use frequency after controlling sex, age, and total intracranial volume. The Benjamini-Hochberg False Discovery Rate method was used for multiple comparison correction of several univariate correlation analyses with a significance level of P < 0.05 [29].

## Results

**Participants and measurements**

A series of self-defined items fully depicted the AICA usage characteristics in the current young adult sample. In the context of first time AICA usage, most participants (58.5%) started to have AICA use within about one year after the release of ChatGPT, and almost no one started using AICA when they were under investigation (0.5%; Fig. S1). All participants had used AICA recently (within the last month), with 48.6% reporting AICA use within the last day and 38.3% reporting AICA use within the last week (Fig. S2). Regarding general AICA use frequency, the majority of participants (61.3%) used AICA frequently throughout the week (at least 4 days a week; Fig. S3). For specific AICA use frequency, the majority of participants (82.5%) reported that they frequently used AICA to meet their functional needs (Fig. S4); however, a smaller number of participants (6.8%) reported that they frequently used AICA to meet their socio-affective needs (Fig. S5). 37.5% of them said they use AICA about an hour or more each day on average (Fig. S6). Most people (59.9%) reported the frequent use of text interaction/creation mode (Fig. S7), but few people reported the frequent use of voice interaction/creation mode (5%; Fig. S8), image interaction/creation mode (9.1%; Fig. S9), and video interaction/creation mode (0.5%; Fig. S10). For the usage rate of AICA products, the vast majority of respondents reported having used DeepSeek (98.2%), Doubao (90.5%), ChatGPT (78.4%), Kimi (73.0%), and ERNIE Bot (58.1%; Fig. S11); however, DeepSeek (44.6%), Doubao (33.3%), and ChatGPT (12.2%) remained the most commonly used AICAs (Fig. S12). For the specific purpose of AICA use, almost all participants had used AICA for academic support (98.2%) and research assistance (90.5%; Fig. S13). Additionally, 22.1% participants were a member (paying a monthly subscription fee) for at least one AICA product (Fig. S14). Finally, the majority of study participants (86.5%) reported to be satisfied with their current AICAs (Fig. S15).

**Coactivation networks of identified brain regions in the regional morphology analyses**

The left dlPFC (-20, 66, -4) associated with general AICA use frequency revealed in the VBM analysis was coactivated with the left middle frontal gyrus (Fig. 3a); the left CAL (-2, -81, 9) associated with general AICA use frequency revealed in the VBM analysis was coactivated with the left supplementary motor area, left inferior frontal gyrus, right inferior occipital gyrus, and left thalamus (Fig. 3b); the left STG (-44, 2, -12) associated with socio-emotional AICA use frequency revealed in the VBM analysis was coactivated with the bilateral insula, right supplementary motor area, right thalamus, and left pallidum (Fig. 3c); the left amygdala (-18, -3, -14) associated with socio-emotional AICA use frequency revealed in the VBM analysis was coactivated with the left insula, bilateral fusiform gyrus, right inferior occipital gyrus, left precentral gyrus, and left superior frontal gyrus (Fig. 3d). A full report of MACM results could be found in the Table. S4.

**Behavioral characterization of identified brain regions and coactivation networks**

The decoding results of regional morphology showed that: 1) brain regions showing GMV

changes associated general AICA use frequency were mainly related with high-order cognition functions (e.g., rules, reasoning, executive function, decision-making) and self-reflection (e.g., default mode, reference); 2) brain regions showing GMV changes associated socio-emotional AICA use frequency were primarily correlated with social and affective processing (e.g., happy faces, neutral faces, pleasant, happy, fear; Fig. 2). The decoding results of coactivated network of regional morphology revealed that: 1) the coactivation patterns of left dlPFC observed in VBM analysis were correlated with high-order cognition function (Fig. 3a), keeping consistent with decoding of regional morphology; the coactivation pattern of left CAL was involved with visual attention processing (Fig. 3b); the coactivation pattern of left STG was linked with auditory processing and social processing (Fig. 3c); the coactivation pattern of left amygdala was associated with social and affective processing (Fig. 3d). These finding keep largely consistent with the behavioral decoding of regional morphology results.

# Supplementary figures and tables

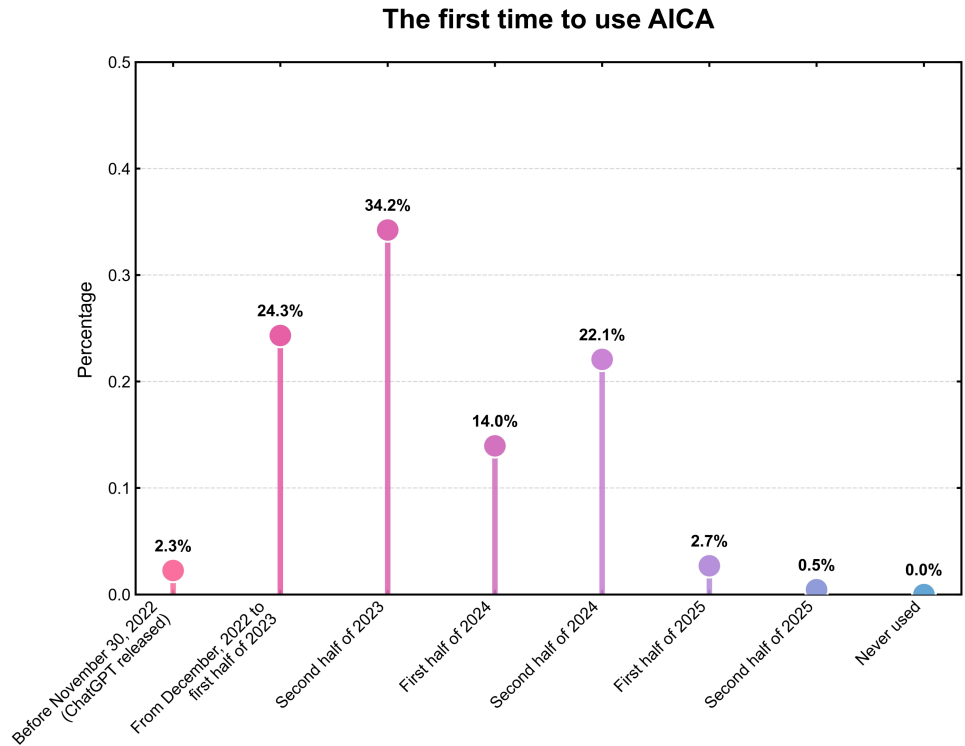

Fig. S1. The frequency distribution statistics regarding the first time of AICA usage. Abbreviation: AICA = Artificial Intelligence-driven Conversational Agent.

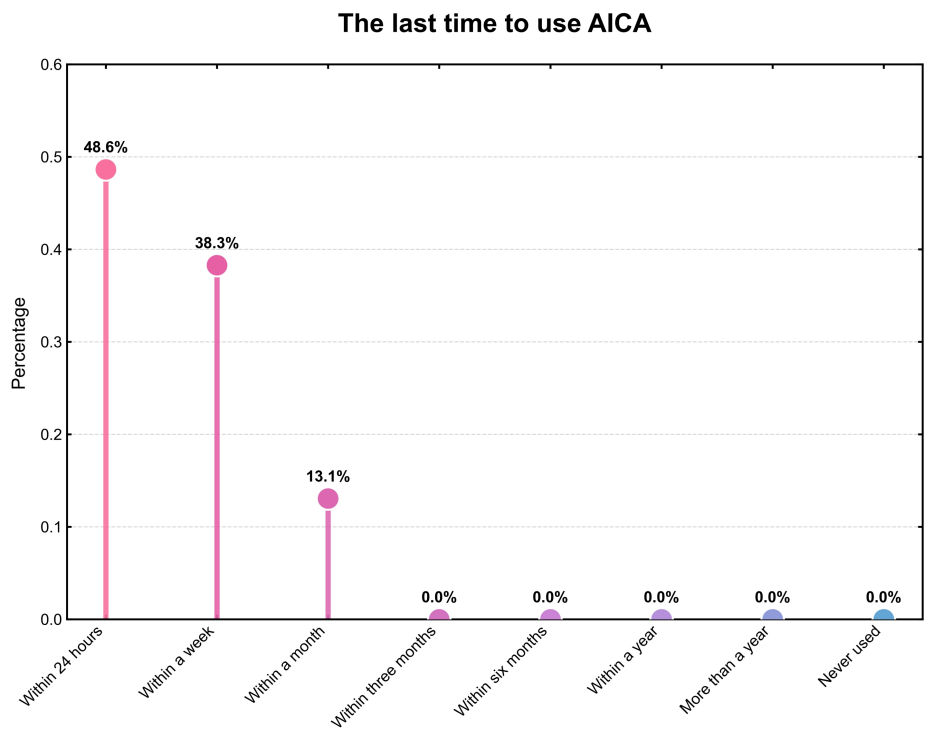

Fig. S2. The frequency distribution statistics regarding the time of last use of AICA. Abbreviation: AICA = Artificial Intelligence-driven Conversational Agent.

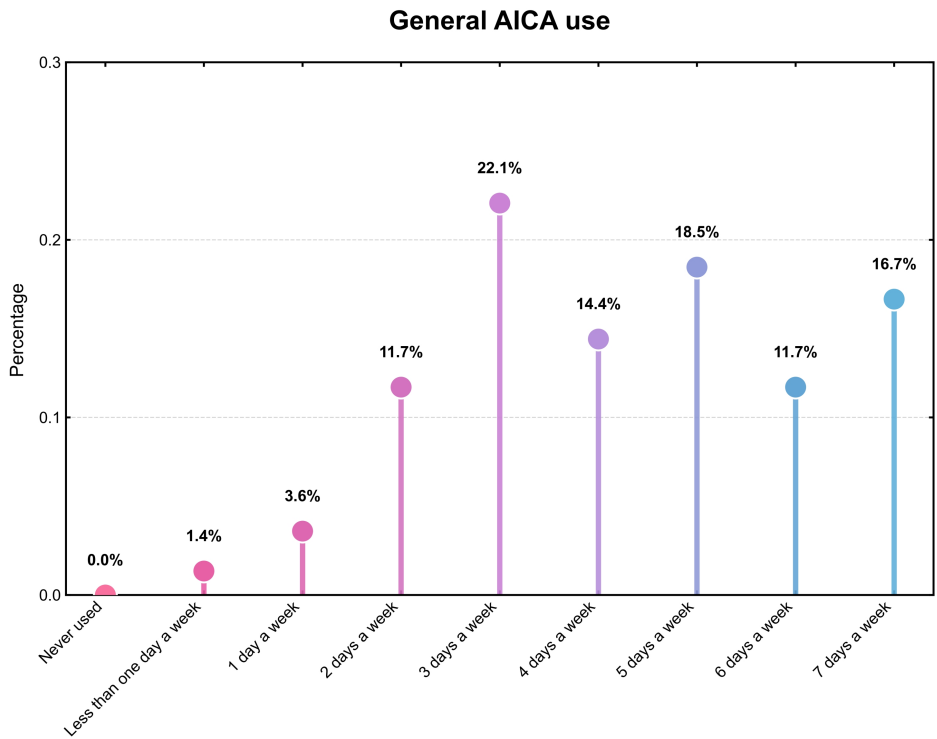

Fig. S3. The frequency distribution statistics regarding the general use of AICA. Abbreviation: AICA = Artificial Intelligence-driven Conversational Agent.

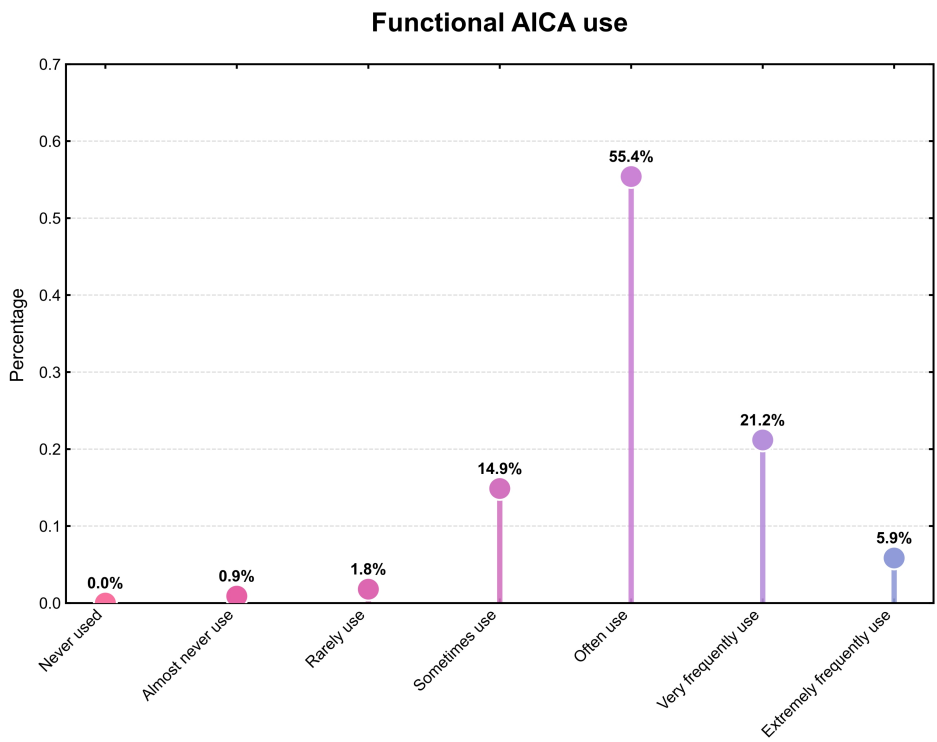

Fig. S4. The frequency distribution statistics regarding the functional use of AICA. Abbreviation: AICA = Artificial Intelligence-driven Conversational Agent.

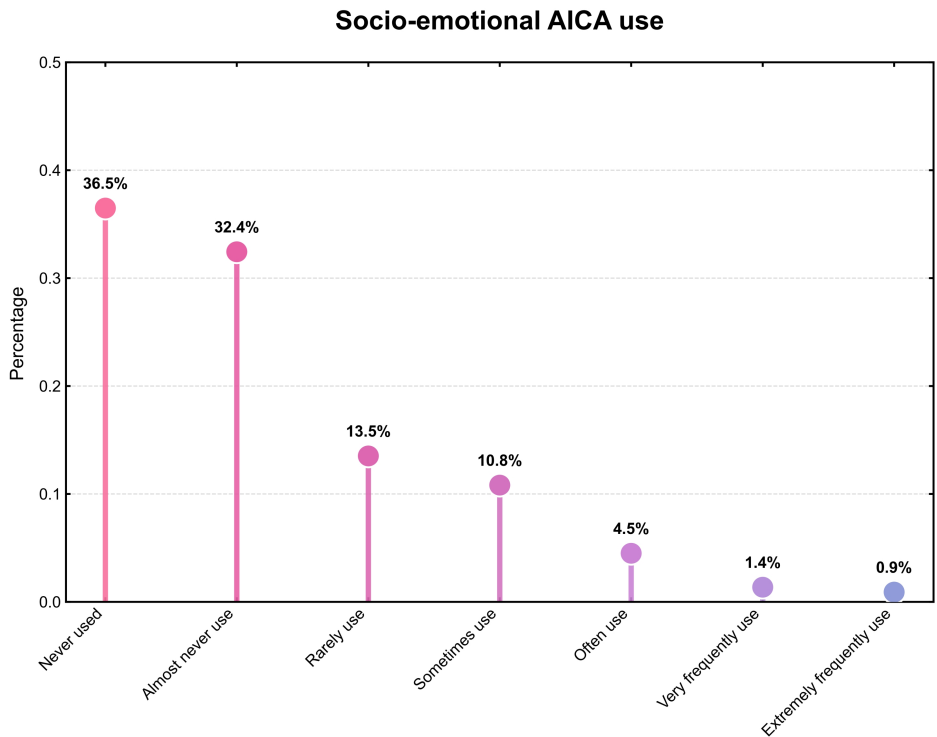

Fig. S5. The frequency distribution statistics regarding the socio-emotional use of AICA. Abbreviation: AICA = Artificial Intelligence-driven Conversational Agent.

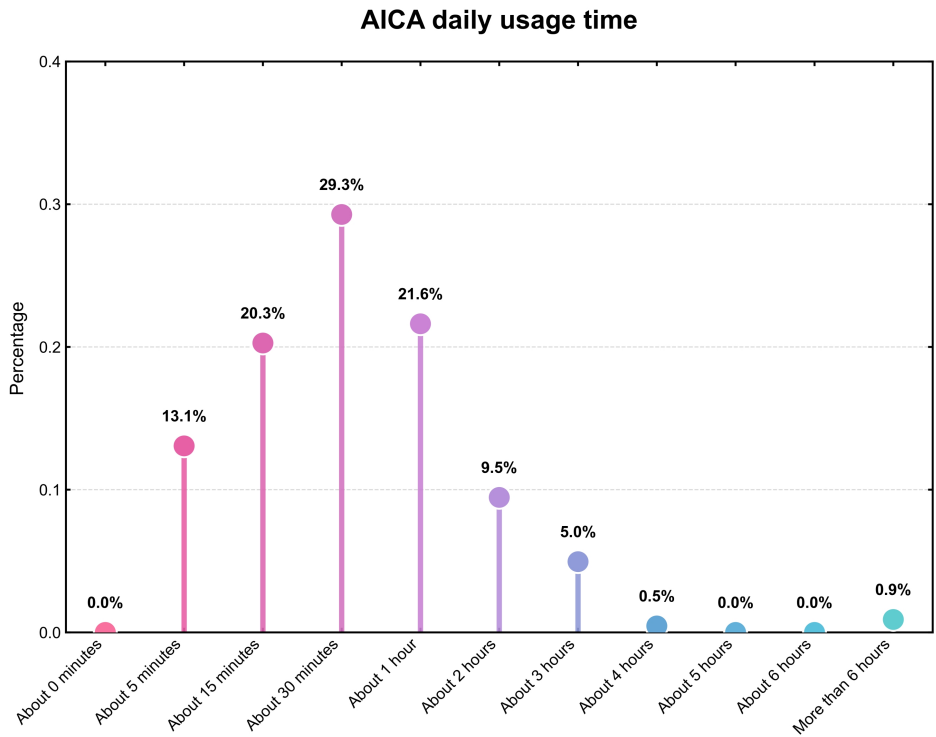

Fig. S6. The frequency distribution statistics regarding the general AICA daily usage time. Abbreviation: AICA = Artificial Intelligence-driven Conversational Agent.

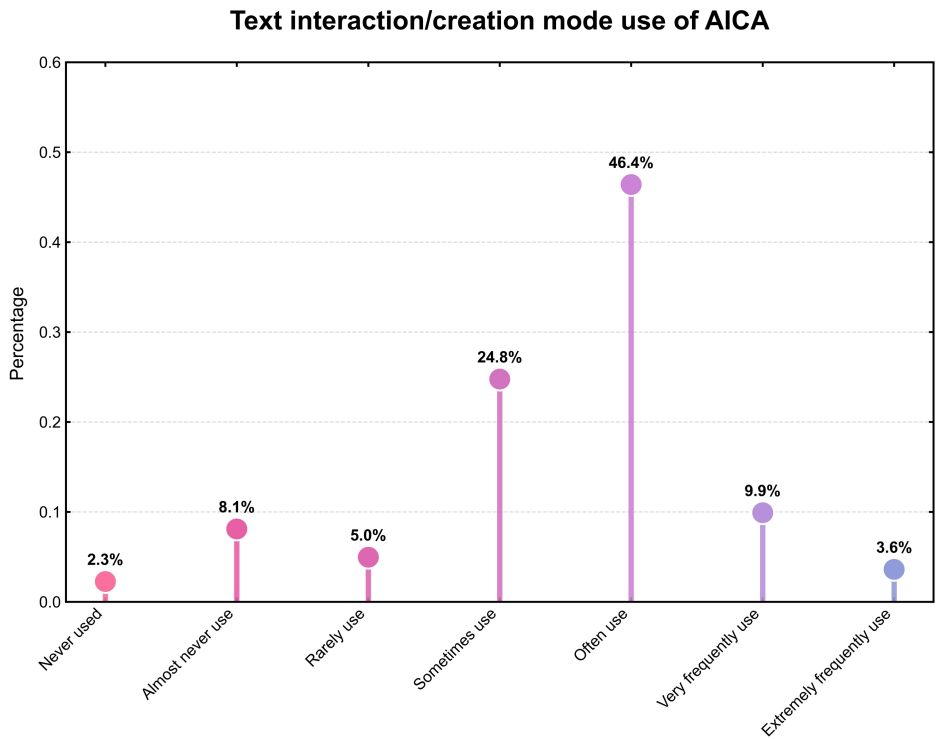

Fig. S7. The frequency distribution statistics regarding the text interaction/creation mode use of AICA. Abbreviation: AICA = Artificial Intelligence-driven Conversational Agent.

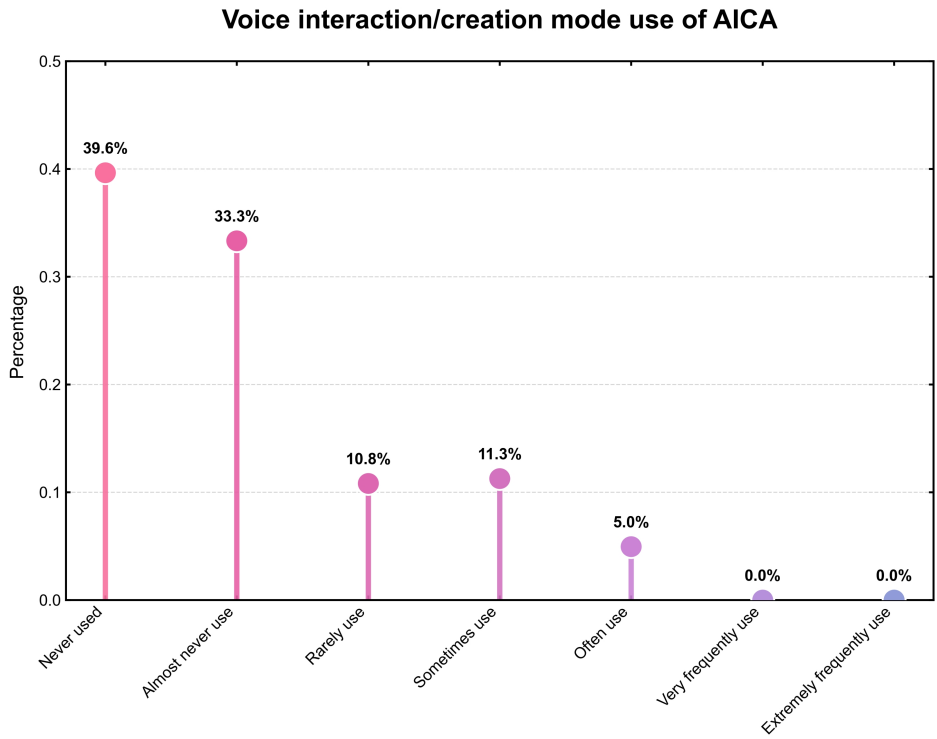

Fig. S8. The frequency distribution statistics regarding the voice interaction/creation mode use of AICA. Abbreviation: AICA = Artificial Intelligence-driven Conversational Agent.

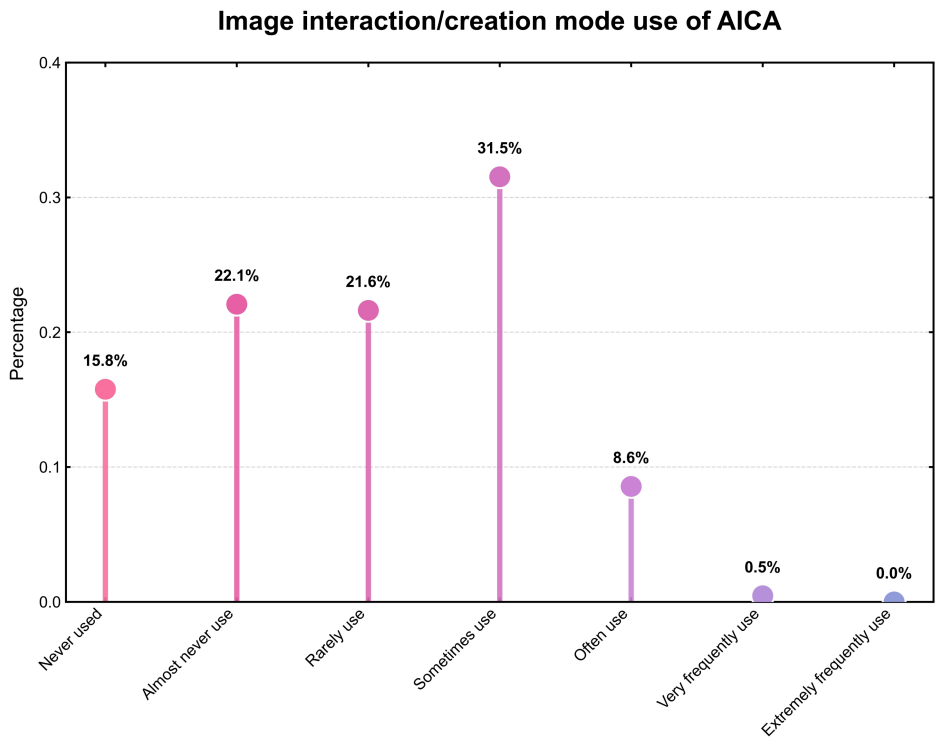

Fig. S9. The frequency distribution statistics regarding the image interaction/creation mode use of AICA. Abbreviation: AICA = Artificial Intelligence-driven Conversational Agent.

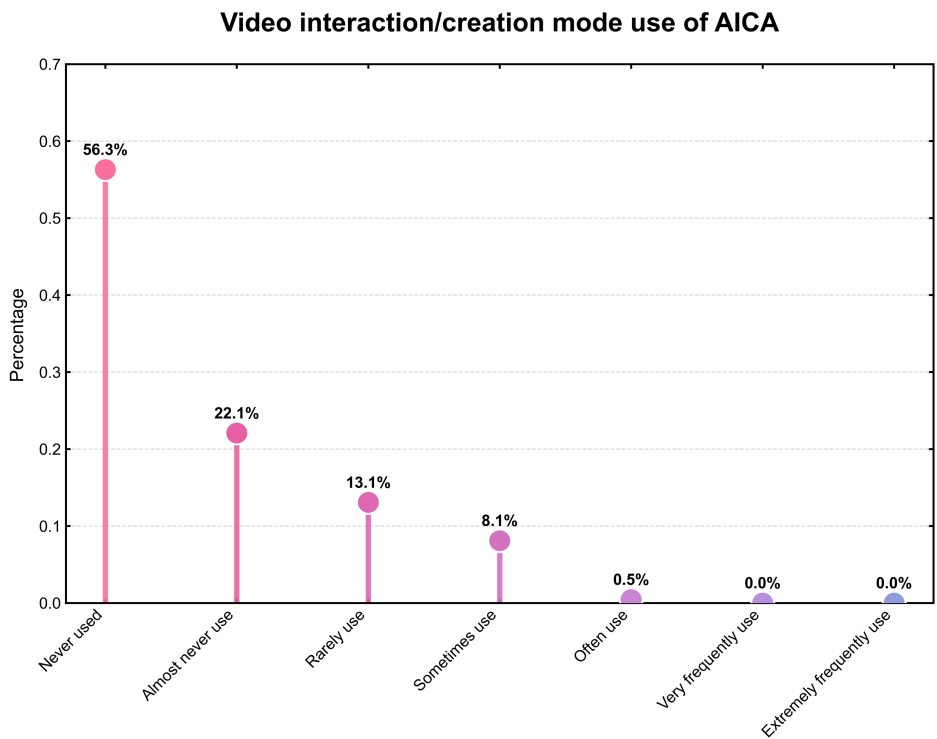

Fig. S10. The frequency distribution statistics regarding the video interaction/creation mode use of AICA. Abbreviation: AICA = Artificial Intelligence-driven Conversational Agent.

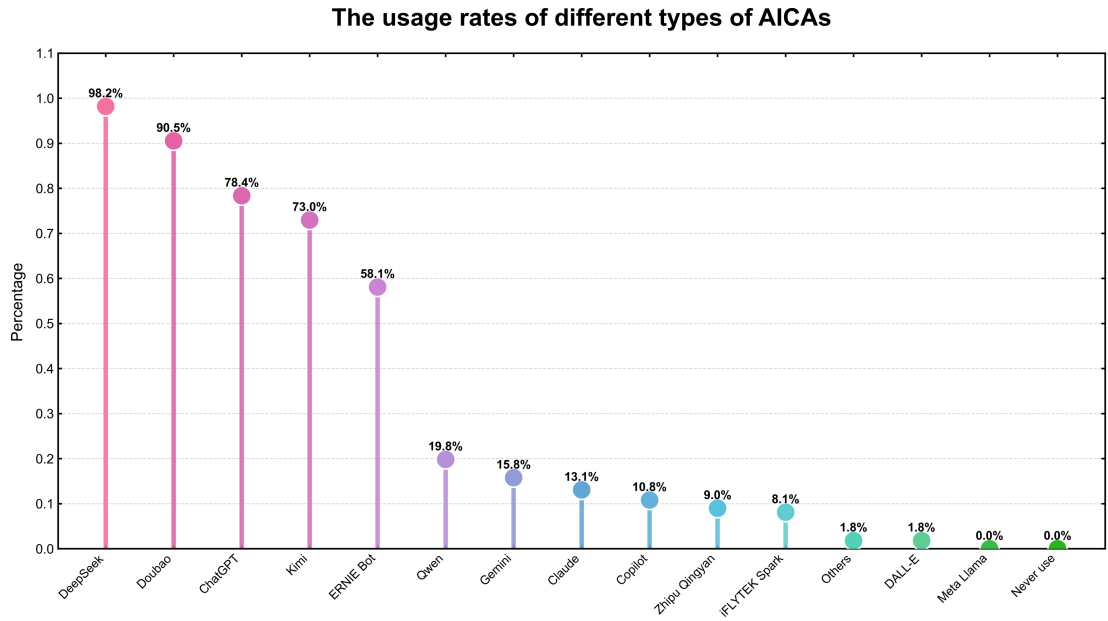

Fig. S11. The frequency distribution statistics regarding the usage rates of different types of AICAs. Abbreviation: AICAs = Artificial Intelligence-driven Conversational Agents.

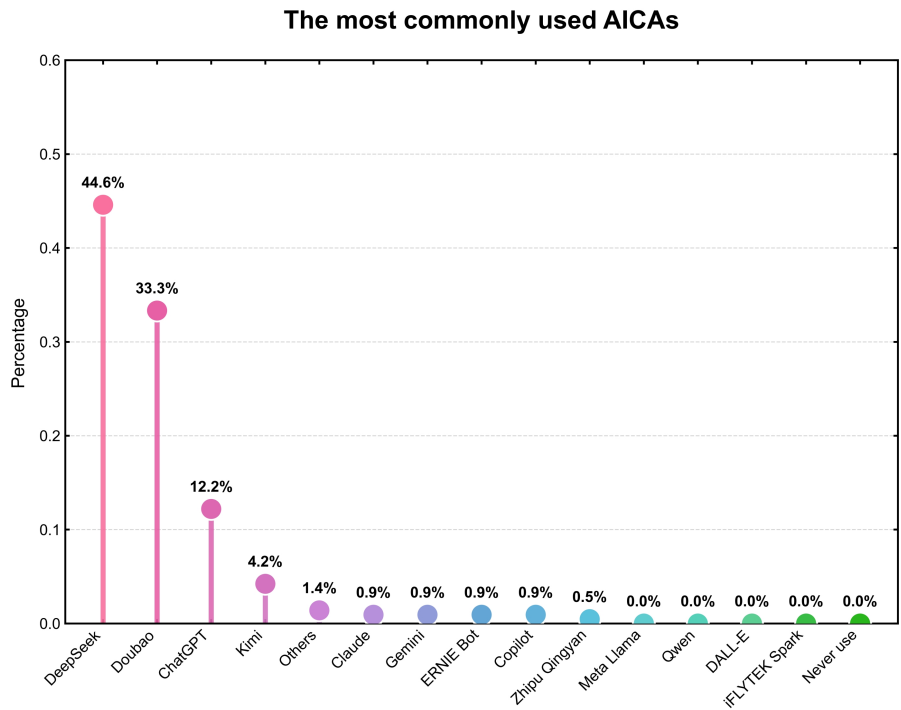

Fig. S12. The frequency distribution statistics regarding the self-reported most commonly used AICAs. Abbreviation: AICAs = Artificial Intelligence-driven Conversational Agents.

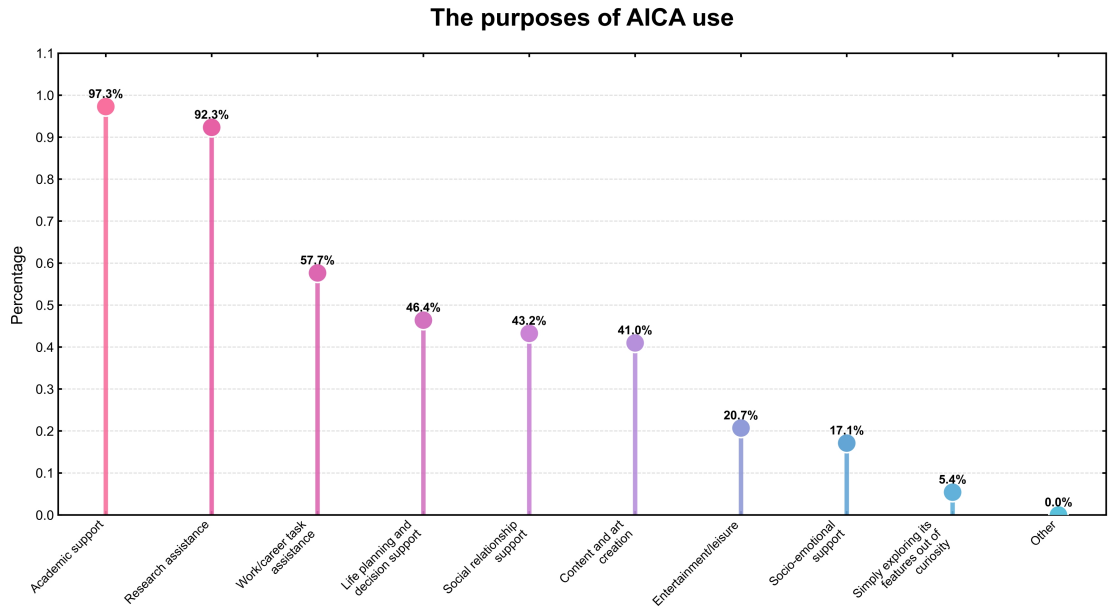

Fig. S13. The frequency distribution statistics regarding the use purpose of AICA. Abbreviation: AICA = Artificial Intelligence-driven Conversational Agent.

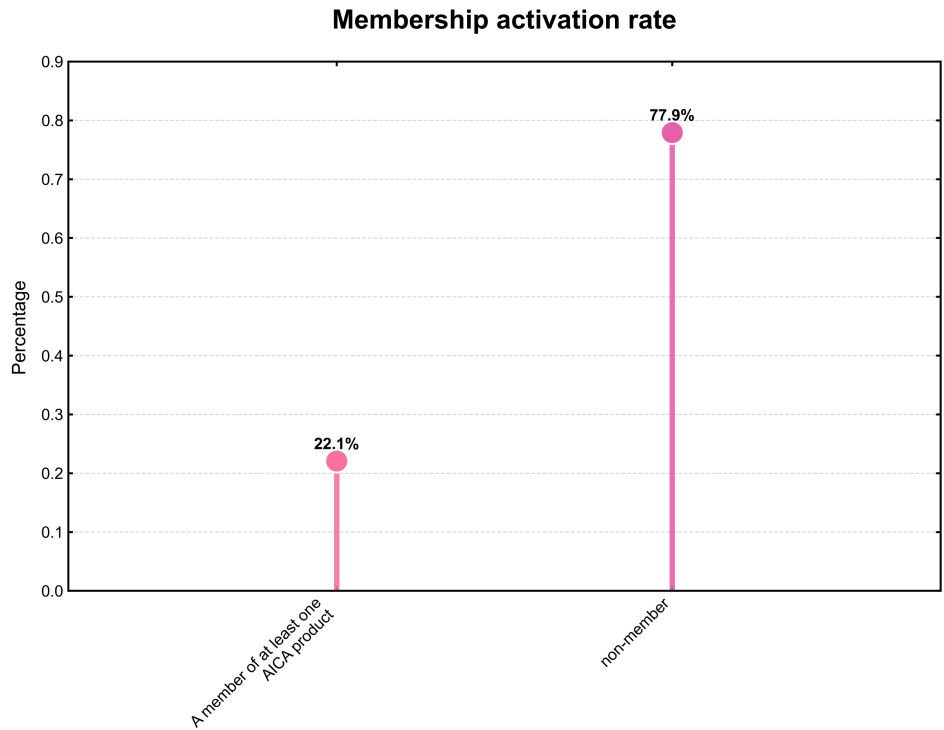

Fig. S14. The membership activation rate of AICA. Abbreviation: AICA = Artificial Intelligence-driven Conversational Agent.

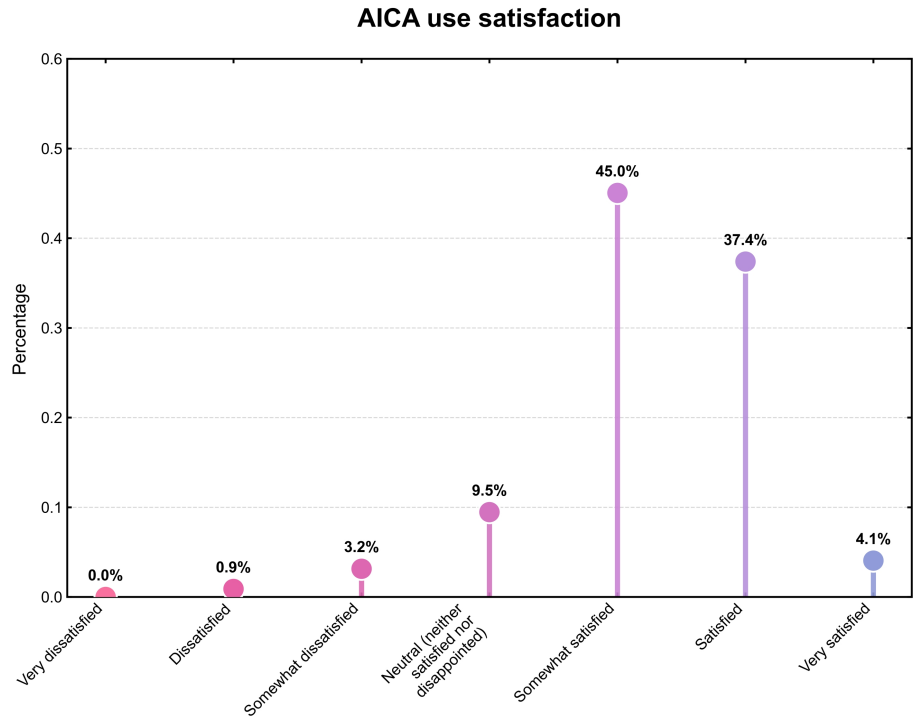

Fig. S15. The frequency distribution statistics regarding the use satisfaction of AICA. Abbreviation: AICA = Artificial Intelligence-driven Conversational Agent.

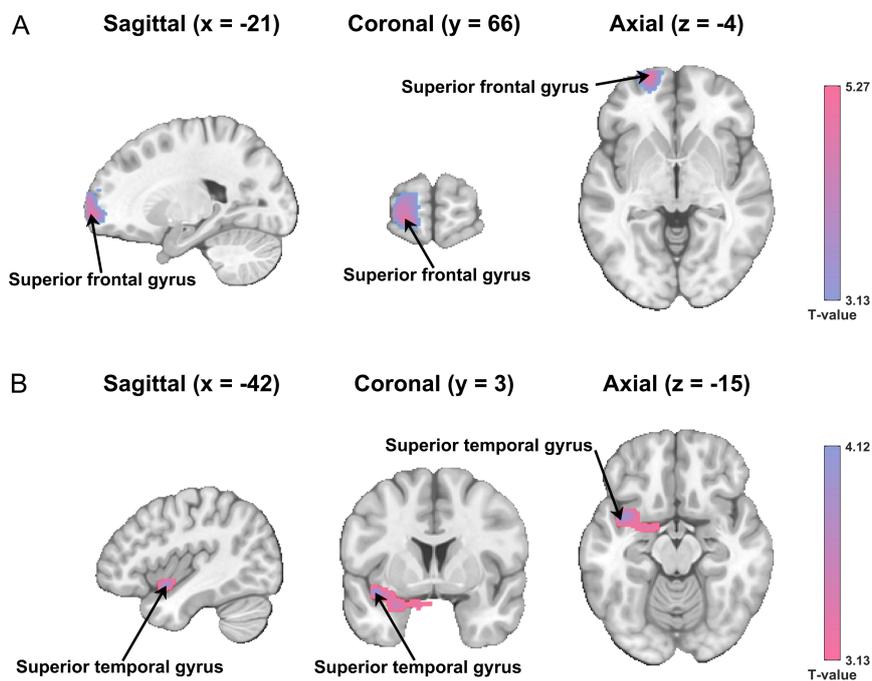

Fig. S16. The brain regions showing morphological alterations correlated with AICA use frequency based on sample containing participants reporting no AICA use within one month (n = 238). A) The brain map depicts a positive association between the frequency of general AICA use and GMV. B). The brain map presents a negative association between the frequency of socio-emotional AICA use and GMV.

Table S1. Descriptive statistics for academic performance and mental health assessment

| Variable | Overall | | Female | | Male | |
|---|---|---|---|---|---|---|
| | M (SD) | Range | M (SD) | Range | M (SD) | Range |
| GPA | 3.50 (0.41) | 2-4 | 3.56 (0.39) | 2-4 | 3.42 (0.43) | 2.11-4 |
| General mental health | 5.04 (1.35) | 2-7 | 5.00 (1.35) | 2-7 | 5.08 (1.36) | 2-7 |
| Loneliness | 2.24 (0.88) | 1-4 | 2.20 (0.93) | 1-4 | 2.29 (0.84) | 1-4 |
| Depression | 3.43 (1.98) | 1-8 | 3.44 (2.04) | 1-8 | 3.42 (1.93) | 1-8 |
| Social anxiety | 21.57 (8.22) | 10-44 | 21.31 (7.98) | 10-44 | 21.87 (8.53) | 10-42 |

Note: GPA = grade-point average, M (SD) = mean (standard deviation).

Table S2. Brain regions associated with AICA use

| Morphological indicator | Behavioral variable | Direction | k (voxels) | $p_{FWE}$ | T | Peak coordinates (x, y, z) | Side | Brain region |
|---|---|---|---|---|---|---|---|---|
| VBM | General AICA use | Positive | 1339 | 0.009 | 5.27 | -20, 66, -4 | Left | Superior frontal gyrus, dorsolateral |
| | | | | | 4.13 | -21, 69, 9 | Left | Superior frontal gyrus, dorsolateral |
| | | | 1676 | 0.003 | 4.28 | -2, -81, 9 | Left | Calcarine fissure and surrounding cortex |
| | | | | | 4.19 | 9, -86, 20 | Right | Cuneus |
| | Socio-emotional AICA use | Negative | 1477 | 0.006 | 4.35 | -44, 2, -12 | Left | Superior temporal gyrus |
| | | | | | 3.83 | -18, -3, -14 | Left | Amygdala |
| | | | | | 3.68 | -28, 6, -21 | Left | Temporal pole: superior temporal gyrus |

Note: AICA = Artificial Intelligence-driven Conversational Agent, VBM = voxel-based morphology, k = cluster size, $p_{FWE}$ = p value after cluster-level FWE correction, T = peak intensity. All coordinates are reported in Montreal Neurological Institute space. The voxel-level threshold is set at p < 0.001, uncorrected; and the cluster-level threshold is set at p < 0.05, FWE-corrected.

Table S3. Brain regions associated with AICA use in the sample containing participants reporting no AICA use within one month (n = 238)

| Morphological indicator | Behavioral variable | Direction | k (voxels) | $p_{FWE}$ | T | Peak coordinates (x, y, z) | Side | Brain region |
|---|---|---|---|---|---|---|---|---|
| VBM | General AICA use | Positive | 1277 | 0.013 | 5.27 | -21, 66, -4 | Left | Superior frontal gyrus, dorsolateral |
| | | | | | 4.49 | -22, 66, 4 | Left | Superior frontal gyrus, dorsolateral |
| | | | | | 3.31 | -20, 60, 16 | Left | Superior frontal gyrus, dorsolateral |

| | | | k | $p_{FWE}$ | T | | | |
|---|---|---|---|---|---|---|---|---|
| Socio-emotional AICA use | Negative | 975 | 0.037 | 4.12 | -42, 3, -14 | Left | Superior temporal gyrus, amygdala |
| | | | | 3.90 | -36, 8, -16 | Left | Insula |
| | | | | 3.65 | -28, 4, -22 | Left | Temporal pole: superior temporal gyrus |

Note: AICA = artificial intelligence-driven conversational agent, VBM = voxel-based morphology, k = cluster size, $p_{FWE}$ = p value after cluster-level FWE correction, T = peak intensity. All coordinates are reported in MNI space. The voxel-level threshold is set at p < 0.001, uncorrected; and the cluster-level threshold is set at p < 0.05, FWE-corrected.

Table S4. The coactivation patterns revealed by meta-analytical coactivation modeling

| Seed region (radius, x, y, z) | Volume (mm³) | Side | Brain region | Coordinates (x, y, z) | ALE value | P | Z |
|---|---|---|---|---|---|---|---|
| Superior frontal gyrus (8, -20, 66, -4) | 4984 | Left | Superior frontal gyrus, dorsolateral | -20, 62, -2 | 0.086 | 2.21E-33 | 11.984 |
| | 1528 | Left | Middle frontal gyrus | -48, 14, 36 | 0.032 | 1.44E-09 | 5.938 |
| Calcarine fissure (8, -2, -81, 9) | 31744 | Left | Calcarine fissure and surrounding cortex | -2, -82, 8 | 0.313 | 0 | 24.080 |
| | | Right | Lingual gyrus | 24, -62, -10 | 0.045 | 1.53E-07 | 5.120 |
| | | Left | Lingual gyrus | -12, -76, -12 | 0.044 | 3.70E-07 | 4.951 |
| | | Right | Calcarine fissure and surrounding cortex | 16, -60, 6 | 0.041 | 1.58E-06 | 4.660 |
| | | Right | Cuneus | 16, -94, 10 | 0.039 | 5.03E-06 | 4.416 |
| | | Right | Lingual gyrus | 18, -58, -4 | 0.038 | 6.62E-06 | 4.356 |
| | | Left | Inferior occipital gyrus | -36, -82, -8 | 0.036 | 1.52E-05 | 4.170 |
| | | Right | Lingual gyrus | 12, -76, -10 | 0.035 | 2.93E-05 | 4.018 |
| | | Left | Inferior occipital gyrus | -24, -88, -6 | 0.034 | 4.13E-05 | 3.937 |
| | | Right | Fusiform gyrus | 26, -74, -14 | 0.033 | 7.09E-05 | 3.805 |
| | | Right | Lingual gyrus | 24, -80, -10 | 0.030 | 3.03E-04 | 3.429 |
| | 6480 | Left | Supplementary motor area | -2, 16, 44 | 0.064 | 2.57E-12 | 6.902 |
| | | Left | Supplementary motor area | -2, 4, 60 | 0.044 | 3.40E-07 | 4.967 |
| | | Right | Middle cingulate & paracingulate gyri | 2, 26, 32 | 0.037 | 1.09E-05 | 4.247 |
| | | Left | Superior frontal gyrus, medial | 0, 26, 36 | 0.037 | 1.16E-05 | 4.232 |
| | 4728 | Left | Inferior frontal gyrus, triangular part | -44, 22, 22 | 0.051 | 6.99E-09 | 5.674 |
| | | Left | Inferior frontal gyrus, triangular part | -52, 22, 16 | 0.042 | 1.13E-06 | 4.728 |
| | | Left | Precentral gyrus | -44, 10, 34 | 0.040 | 2.74E-06 | 4.545 |
| | | Left | Inferior frontal gyrus, triangular part | -46, 14, 24 | 0.038 | 5.70E-06 | 4.389 |
| | | Left | Inferior frontal gyrus, opercular part | -48, 12, 10 | 0.034 | 5.11E-05 | 3.885 |
| | 3424 | Right | Inferior occipital gyrus | 44, -76, -8 | 0.042 | 1.06E-06 | 4.742 |
| | | Right | Inferior occipital gyrus | 44, -72, -12 | 0.041 | 1.23E-06 | 4.711 |
| | | Right | Inferior temporal gyrus | 48, -56, -12 | 0.037 | 1.00E-05 | 4.264 |
| | | Right | Middle temporal gyrus | 48, -74, 2 | 0.035 | 3.57E-05 | 3.972 |
| | | Right | Inferior occipital gyrus | 38, -84, -4 | 0.034 | 5.85E-05 | 3.852 |
| | | Right | Inferior occipital gyrus | 32, -88, -4 | 0.033 | 6.61E-05 | 3.822 |
| | | Right | Lobule VI of cerebellar hemisphere | 36, -66, -22 | 0.029 | 4.05E-04 | 3.350 |
| | 3120 | Left | Mediodorsal lateral parvocellular thalamus | -10, -18, 8 | 0.043 | 5.22E-07 | 4.883 |
| | | Left | Red nucleus | 2, -24, -8 | 0.030 | 3.69E-04 | 3.375 |
| Superior temporal gyrus (8, -44, 2, -12) | 25552 | Left | Superior temporal gyrus | -42, 0, -12 | 0.208 | 0 | 18.740 |
| | | Left | Insula | -34, 24, 4 | 0.045 | 1.37E-09 | 5.947 |
| | | Left | Superior temporal gyrus | -56, -16, 2 | 0.041 | 1.08E-08 | 5.599 |
| | | Left | Middle temporal gyrus | -60, -34, 6 | 0.040 | 2.95E-08 | 5.422 |
| | | Left | Supramarginal gyrus | -56, -22, 14 | 0.039 | 5.65E-08 | 5.304 |
| | | Left | Superior temporal gyrus | -62, -30, 12 | 0.038 | 7.01E-08 | 5.265 |
| | | Left | Supramarginal gyrus | -60, -24, 32 | 0.035 | 4.94E-07 | 4.894 |
| | | Left | Insula | -34, 8, 6 | 0.026 | 7.46E-05 | 3.792 |
| | | Left | Superior temporal gyrus | -42, -26, 4 | 0.024 | 1.98E-04 | 3.542 |
| | | Left | Supramarginal gyrus | -52, -28, 34 | 0.024 | 2.00E-04 | 3.540 |
| | 12400 | Right | Insula | 40, 4, -14 | 0.047 | 3.17E-10 | 6.182 |
| | | Right | Superior temporal gyrus | 58, -4, -2 | 0.042 | 1.03E-08 | 5.607 |
| | | Right | Superior temporal gyrus | 58, -24, 2 | 0.041 | 1.66E-08 | 5.524 |
| | | Right | Superior temporal gyrus | 54, -8, 4 | 0.036 | 3.05E-07 | 4.988 |
| | | Right | Inferior frontal gyrus, opercular part | 50, 16, 2 | 0.030 | 6.62E-06 | 4.356 |
| | | Right | Insula | 44, 14, -10 | 0.030 | 7.89E-06 | 4.318 |
| | | Right | Middle temporal gyrus | 56, -34, 4 | 0.029 | 1.73E-05 | 4.141 |
| | | Right | Temporal pole: middle temporal gyrus | 52, 4, -18 | 0.022 | 4.03E-04 | 3.351 |
| | 4376 | Right | Supplementary motor area | 4, 4, 46 | 0.039 | 6.37E-08 | 5.282 |
| | | Left | Supplementary motor area | -6, 14, 52 | 0.036 | 2.55E-07 | 5.023 |
| | | Left | Middle cingulate & paracingulate gyri | -4, 14, 42 | 0.032 | 2.99E-06 | 4.527 |
| | | Left | Middle cingulate & paracingulate gyri | 0, 18, 32 | 0.026 | 6.91E-05 | 3.812 |
| | 2296 | Right | Ventral lateral thalamus | 14, -12, 2 | 0.039 | 4.08E-08 | 5.363 |
| | | Right | Red nucleus | 6, -20, -8 | 0.025 | 1.40E-04 | 3.633 |
| | | 1896 | Right | Insula | 34, 22, 2 | 0.040 | 3.04E-08 | 5.417 |
| | | 1616 | Left | Lenticular nucleus, Pallidum | -20, 2, 0 | 0.029 | 1.49E-05 | 4.175 |
| | | | Left | Ventral lateral thalamus | -16, -10, 2 | 0.026 | 5.98E-05 | 3.847 |
| Amygdala (6, -18, -3, -14) | 35456 | Left | Amygdala | -18, -4, -14 | 0.473 | 0 | 33.823 |
| | | Right | Hippocampus | 20, -4, -14 | 0.209 | 0 | 17.152 |
| | | Right | Ventral lateral thalamus | 6, -10, 6 | 0.056 | 3.23E-10 | 6.179 |
| | | Left | Lenticular nucleus, Pallidum | -18, 6, 2 | 0.047 | 4.12E-08 | 5.362 |
| | | Right | Nucleus accumbens | 10, 10, -6 | 0.043 | 4.21E-07 | 4.925 |
| | | Right | Lenticular nucleus, Putamen | 20, 10, -2 | 0.038 | 6.27E-06 | 4.368 |
| | | Right | Lateral geniculate thalamus | 22, -22, -8 | 0.037 | 1.02E-05 | 4.261 |
| | | Right | Caudate nucleus | 12, 4, 12 | 0.036 | 1.28E-05 | 4.210 |
| | | Left | Thalamus | 2, -26, 2 | 0.035 | 2.03E-05 | 4.104 |
| | | Right | Thalamus | 6, -24, -2 | 0.035 | 2.24E-05 | 4.081 |
| | | Left | Parahippocampal gyrus | -22, -26, -16 | 0.032 | 9.93E-05 | 3.721 |
| | | Left | Thalamus | -6, -30, -8 | 0.031 | 1.42E-04 | 3.630 |
| | 5168 | Left | Insula | -34, 24, 0 | 0.043 | 4.26E-07 | 4.923 |
| | | Left | Inferior frontal gyrus, triangular part | -36, 28, -2 | 0.043 | 4.37E-07 | 4.918 |
| | | Left | Inferior frontal gyrus, pars orbitalis | -46, 24, -8 | 0.037 | 6.61E-06 | 4.356 |
| | | Left | Insula | -36, 12, -4 | 0.035 | 1.96E-05 | 4.113 |
| | | Left | Posterior orbital gyrus | -34, 24, -14 | 0.031 | 1.38E-04 | 3.637 |
| | | Left | Inferior frontal gyrus, pars orbitalis | -28, 26, -10 | 0.028 | 4.20E-04 | 3.339 |
| | 4312 | Left | Fusiform gyrus | -42, -58, -18 | 0.060 | 2.18E-11 | 6.591 |
| | | Left | Inferior occipital gyrus | -40, -74, -10 | 0.038 | 5.20E-06 | 4.409 |
| | 2664 | Right | Inferior occipital gyrus | 50, -74, -6 | 0.047 | 3.84E-08 | 5.374 |
| | | Right | Inferior occipital gyrus | 40, -72, -8 | 0.036 | 1.62E-05 | 4.155 |
| | 2456 | Right | Fusiform gyrus | 42, -50, -18 | 0.073 | 6.78E-15 | 7.701 |
| | 1808 | Left | Precentral gyrus | -52, 10, 32 | 0.036 | 1.28E-05 | 4.209 |
| | | Left | Inferior frontal gyrus, triangular part | -46, 20, 22 | 0.036 | 1.51E-05 | 4.172 |
| | | Left | Inferior frontal gyrus, triangular part | -44, 32, 14 | 0.032 | 9.36E-05 | 3.736 |
| | | Left | Inferior frontal gyrus, pars orbitalis | -40, 6, 24 | 0.027 | 7.24E-04 | 3.185 |
| | 1648 | Left | Superior frontal gyrus, medial | -2, 60, 18 | 0.045 | 1.56E-07 | 5.117 |

Note: ALE = activation likelihood estimate. All coordinates are reported in MNI space. The voxel-level threshold is set at p < 0.001, uncorrected; and the cluster-level threshold is set at p < 0.05, FWE-corrected.

# Appendix

**Notes: Full description of the items related with AICA use and mental health (The original version was presented in Chinese; here, it has been translated into English.**

- *Items measuring AICA use characteristics*
**1. When was the first time you used AICA?**
1) Before November 30, 2022 (ChatGPT released)
2) From December, 2022 to first half of 2023
3) Second half of 2023
4) First half of 2024
5) Second half of 2024
6) First half of 2025
7) Second half of 2025
8) Never used

**2. When was the last time you used AICA?**
1) Within 24 hours
2) Within a week
3) Within a month
4) Within three months
5) Within six months
6) Within a year
7) More than a year
8) Never used

**3. How often do you use AICA on average per week?**
1) Never used
2) Less than one day a week
3) 1 day a week
4) 2 days a week
5) 3 days a week
6) 4 days a week
7) 5 days a week
8) 6 days a week
9) 7 days a week

**4. To what extent do you use AICA to obtain functional support (such as advice, opinion, and assistance for work, study, and academics)?**
1) Never used
2) Almost never use
3) Rarely use
4) Sometimes use
5) Often use
6) Very frequently use
7) Extremely frequently use

**5. To what extent do you use AICA to obtain socio-emotional support (such as confiding in AI,**

**receiving AI companionship, and developing friendship or romantic relationship with AI)?**
1) Never used
2) Almost never use
3) Rarely use
4) Sometimes use
5) Often use
6) Very frequently use
7) Extremely frequently use

**6. How much time do you spend on it each day on average in recent one semester? (Excluding time spent running in the background)**
1) About 0 minutes (not used or used very rarely)
2) About 5 minutes
3) About 15 minutes
4) About 30 minutes
5) About 1 hour
6) About 2 hours
7) About 3 hours
8) About 4 hours
9) About 5 hours
10) About 6 hours
11) More than 6 hours

**7. How often do you use the text interaction/creation mode of AICA?**
1) Never used
2) Almost never use
3) Rarely use
4) Sometimes use
5) Often use
6) Very frequently use
7) Extremely frequently use

**8. How often do you use the voice interaction/creation mode of AICA?**
1) Never used
2) Almost never use
3) Rarely use
4) Sometimes use
5) Often use
6) Very frequently use
7) Extremely frequently use

**9. How often do you use the image interaction/creation mode of AICA?**
1) Never used
2) Almost never use
3) Rarely use
4) Sometimes use
5) Often use
6) Very frequently use

7) Extremely frequently use
**10. How often do you use the video interaction/creation mode of AICA?**
1) Never used
2) Almost never use
3) Rarely use
4) Sometimes use
5) Often use
6) Very frequently use
7) Extremely frequently use
**11. Which AICAs have you used so far? (Multiple choice)**
1) ChatGPT
2) DeepSeek
3) Kimi
4) Claude
5) Gemini
6) Copilot
7) Meta Llama
8) DALL-E
9) Doubao
10) ERNIE Bot
11) Qwen
12) iFLYTEK Spark
13) Zhipu Qingyan
14) Never use
15) Others
**12. Which AICAs have you used most frequently so far?**
1) ChatGPT
2) DeepSeek
3) Kimi
4) Claude
5) Gemini
6) Copilot
7) Meta Llama
8) DALL-E
9) Doubao
10) ERNIE Bot
11) Qwen
12) iFLYTEK Spark
13) Zhipu Qingyan
14) Never use
15) Others
**13. More specifically, what do you typically use AICA for? (Multiple choice)**
1) Academic support
2) Research assistance

3) Content and art creation
4) Entertainment/leisure
5) Life planning and decision support
6) Work/career task assistance
7) Social relationship support
8) Socio-emotional support
9) Simply exploring its features out of curiosity
10) Other

**14. Do you have a membership for at least one AICA product?**
1) Yes
2) No

**15. What is your overall satisfaction with the AICAs you frequently use?**
1) Very dissatisfied
2) Dissatisfied
3) Somewhat dissatisfied
4) Neutral (neither satisfied nor disappointed)
5) Somewhat satisfied
6) Satisfied
7) Very satisfied

- *Items measuring mental health and academic performance (Q16 measured general status of mental health; Q17 measured loneliness; Q18 measured depression; Total score of Q19-Q28 measured social anxiety; Q29 collected GPA).*

**16. How has your general mental health been over the past month?**
1) Very bad
2) Fairly bad
3) Somewhat bad
4) Neither good nor bad
5) Somewhat good
6) Fairly good
7) Very good

**17. How long in the past month have you felt lonely or isolated?**
1) Never
2) Rarely
3) Sometimes
4) Often

**18. Over the past month, how much have you bothered by feeling sad, down, or uninterested in life?**
0) Not at all
1) 
2) 
3) A little
4) 
5) 
6) Moderately

7)
8)
9) Severely

**19. I have difficulty making eye-contact with others.**
1) Not at all characteristic or true of me
2) Slightly characteristic or true of me
3) Moderately characteristic or true of me
4) Very characteristic or true of me
5) Extremely characteristic or true of me

**20. When mixing socially I am uncomfortable.**
1) Not at all characteristic or true of me
2) Slightly characteristic or true of me
3) Moderately characteristic or true of me
4) Very characteristic or true of me
5) Extremely characteristic or true of me

**21. I feel tense if I am alone with just one other person.**
1) Not at all characteristic or true of me
2) Slightly characteristic or true of me
3) Moderately characteristic or true of me
4) Very characteristic or true of me
5) Extremely characteristic or true of me

**22. I have difficulty talking with other people.**
1) Not at all characteristic or true of me
2) Slightly characteristic or true of me
3) Moderately characteristic or true of me
4) Very characteristic or true of me
5) Extremely characteristic or true of me

**23. I worry about expressing myself in case I appear awkward.**
1) Not at all characteristic or true of me
2) Slightly characteristic or true of me
3) Moderately characteristic or true of me
4) Very characteristic or true of me
5) Extremely characteristic or true of me

**24. I find myself worrying that I won't know what to say in social situations.**
1) Not at all characteristic or true of me
2) Slightly characteristic or true of me
3) Moderately characteristic or true of me
4) Very characteristic or true of me
5) Extremely characteristic or true of me

**25. I am nervous mixing with people I don't know well.**
1) Not at all characteristic or true of me
2) Slightly characteristic or true of me
3) Moderately characteristic or true of me
4) Very characteristic or true of me

5) Extremely characteristic or true of me

**26. I feel I'll say something embarrassing when talking.**
1) Not at all characteristic or true of me
2) Slightly characteristic or true of me
3) Moderately characteristic or true of me
4) Very characteristic or true of me
5) Extremely characteristic or true of me

**27. When mixing in a group I find myself worrying I will be ignored.**
1) Not at all characteristic or true of me
2) Slightly characteristic or true of me
3) Moderately characteristic or true of me
4) Very characteristic or true of me
5) Extremely characteristic or true of me

**28. I am tense mixing in a group.**
1) Not at all characteristic or true of me
2) Slightly characteristic or true of me
3) Moderately characteristic or true of me
4) Very characteristic or true of me
5) Extremely characteristic or true of me

**29. Please fill in your GPA after looking up it in the academic system (rounded to two decimal places).**

______________